\documentclass[aps,amsmath,preprint,superscriptaddress,nofootinbib]{revtex4-1}
\usepackage{graphicx}
\usepackage{bm}
\usepackage{color}
\usepackage{amsmath,amssymb}	
\usepackage{hyperref}
\usepackage{setspace}

\usepackage[english]{babel}
\usepackage[utf8x]{inputenc}
\usepackage[colorinlistoftodos]{todonotes}
\usepackage{slashed}
\usepackage{comment}
\usepackage{ulem}

\usepackage{physics}

\begin{document}

%%%%%%%%%%%%%%%%%%%%%%%%%%%%%%%%%%%%%%%%%%%

\def\a{\alpha}
\def\b{\beta}
\def\c{\varepsilon}
\def\d{\delta}
\def\e{\epsilonsilon}
\def\f{\phi}
\def\g{\gamma}
\def\h{\theta}
\def\k{\kappa}
\def\l{\lambda}
\def\m{\mu}
\def\n{\nu}
\def\p{\psi}
\def\q{\partial}
\def\r{\rho}
\def\s{\sigma}
\def\t{\tau}
\def\u{\upsilon}
\def\v{\varphi}
\def\w{\omega}
\def\x{\xi}
\def\y{\eta}
\def\z{\zeta}
\def\D{\Delta}
\def\G{\Gamma}
\def\H{\Theta}
\def\L{\Lambdabda}
\def\F{\Phi}
\def\P{\Psi}
\def\S{\Sigma}

\def\o{\over}
\def\beq{\begin{eqnarray}}
\def\eeq{\end{eqnarray}}
\newcommand{\gsim}{ \mathop{}_{\textstyle \sim}^{\textstyle >} }
\newcommand{\lsim}{ \mathop{}_{\textstyle \sim}^{\textstyle <} }
\newcommand{\EV}{ {\rm eV} }
\newcommand{\KEV}{ {\rm keV} }
\newcommand{\MEV}{ {\rm MeV} }
\newcommand{\GEV}{ {\rm GeV} }
\newcommand{\TEV}{ {\rm TeV} }
\newcommand{\1}{\mbox{1}\hspace{-0.25em}\mbox{l}}
\newcommand{\headline}[1]{\noindent{\bf #1}}
\def\diag{\mathop{\rm diag}\nolimits}
\def\Spin{\mathop{\rm Spin}}
\def\SO{\mathop{\rm SO}}
\def\O{\mathop{\rm O}}
\def\SU{\mathop{\rm SU}}
\def\U{\mathop{\rm U}}
\def\Sp{\mathop{\rm Sp}}
\def\SL{\mathop{\rm SL}}
\def\tr{\mathop{\rm tr}}
\def\mpl{M_{PL}}

\def\IJMP{Int.~J.~Mod.~Phys. }
\def\MPL{Mod.~Phys.~Lett. }
\def\NP{Nucl.~Phys. }
\def\PL{Phys.~Lett. }
\def\PR{Phys.~Rev. }
\def\PRL{Phys.~Rev.~Lett. }
\def\PTP{Prog.~Theor.~Phys. }
\def\ZP{Z.~Phys. }

\def\dd{\mathrm{d}}
\def\ff{\mathrm{f}}
\def\BH{{\rm BH}}
\def\inf{{\rm inf}}
\def\ev{{\rm evap}}
\def\eq{{\rm eq}}
\def\SM{{\rm sm}}
\def\Mpl{M_{\rm PL}}
\def\GeV{{\rm GeV}}
\newcommand{\Red}[1]{\textcolor{red}{#1}}

% To increase line space in footnote.
%

%%%%%%%%%%%%%%%%%%%%%%%%%%%%%%%%%%
%%%%%%%%%%% Title page %%%%%%%%%%%
%%%%%%%%%%%%%%%%%%%%%%%%%%%%%%%%%%

\preprint{RUP-21-17}
\bigskip

\title{Gauge Kinetic Mixing and Dark Topological Defects}

\author{Takashi Hiramatsu}
\email[e-mail: ]{hiramatz@icrr.u-tokyo.ac.jp}
\affiliation{Department of Physics, Rikkyo University, Toshima, Tokyo 171-8501, Japan}
\affiliation{ICRR, The University of Tokyo, Kashiwa, Chiba 277-8582, Japan}
\author{Masahiro Ibe}
\email[e-mail: ]{ibe@icrr.u-tokyo.ac.jp}
\affiliation{ICRR, The University of Tokyo, Kashiwa, Chiba 277-8582, Japan}
\affiliation{Kavli IPMU (WPI), UTIAS, The University of Tokyo, Kashiwa, Chiba 277-8583, Japan}
\author{Motoo Suzuki}
\email[e-mail: ]{m0t@icrr.u-tokyo.ac.jp}
\affiliation{Tsung-Dao Lee Institute and School of Physics and Astronomy, Shanghai
Jiao Tong University, 800 Dongchuan Road, Shanghai, 200240 China}
\author{Soma Yamaguchi}
\email[e-mail: ]{soma@icrr.u-tokyo.ac.jp}
\affiliation{Morpho, Inc., Chiyoda, Tokyo 101-0065, Japan}
\date{\today}

\begin{abstract}
We discuss how the topological defects in the dark sector affect the Standard Model sector when the dark photon has a kinetic mixing with the QED photon. In particular, we consider the dark photon appearing in the successive gauge symmetry breaking, $\mathrm{SU}(2)\to \mathrm{U}(1) \to \mathbb{Z}_2$, where the remaining $\mathbb{Z}_2$ is the center of $\mathrm{SU(2)}$. 
In this model, the monopole is trapped into the cosmic strings and forms the so-called bead solution. As we will discuss, the dark cosmic string induces the QED magnetic flux inside the dark string through the kinetic mixing. The dark monopole, on the other hand, does not induce the QED magnetic flux in the U(1) symmetric phase, even in the presence of the kinetic mixing. Finally, we show that the dark bead solution induces a spherically symmetric QED magnetic flux through the kinetic mixing. The induced flux looks like the QED magnetic monopole viewed from a distance, although QED satisfies the Bianchi identity everywhere, which we call a pseudo magnetic monopole.
\end{abstract}

\maketitle

\section{Introduction}
\label{sec:introduction}

The dark photon, a gauge boson of a new U(1) gauge symmetry, appears in the various context of the extensions of the Standard Model (SM).
In the simplest version, the dark photon couples to the SM sector through the kinetic mixing with the gauge boson of the U(1) in the SM~\cite{Holdom:1985ag}.
Recently, the dark photon has attracted attention for cosmological reasons. For instance, the new U(1) gauge symmetry can be the origin of the stability of dark matter. It is also discussed that the dark matter self-interaction mediated by the sub-GeV dark photon can solve the so-called small scale structure problems~\cite{Spergel:1999mh,Kaplinghat:2015aga,Kamada:2016euw,Tulin:2017ara,Chu:2018fzy,Chu:2019awd}.%
\footnote{See also e.g. Refs~\cite{Hayashi:2020syu,Ebisu:2021bjh}, for recent discussions on the constraints on the dark matter self-interaction cross section using the  Ultra-faint Dwarf Galaxies.}
The dark photon also plays an essential role in transferring excessive entropy in the dark sector to the visible sector in many dark matter models (see e.g. Ref.~\cite{Blennow:2012de,Ibe:2018juk}).
In the wake of the attention, many new experiments are proposed to search for the sub-GeV dark photon  (see Ref.~\cite{Raggi:2015yfk,Bauer:2018onh} for a summary).

A dark photon with a mass would associate with  spontaneously broken U(1) gauge symmetry.%
\footnote{For models with dynamical U(1) breaking, see Refs.~\cite{Co:2016akw,Ibe:2021gil}.}
For several reasons, however, it is more desirable to associate the dark photon with a non-Abelian gauge symmetry.
First, the U(1) gauge theory 
has a Landau pole at some high energy scale, and hence, it
is not ultraviolet (UV) complete. 
The non-Abelian extension renders the model asymptotically free at UV.
The non-Abelian extension is also attractive as it can naturally explain a
tiny kinetic mixing parameter (see e.g., Refs.~\cite{Ibe:2018tex,Ibe:2019ena}).
The tiny kinetic mixing parameter is important for the sub-GeV dark photon to evade all the astrophysical, cosmological, and experimental constraints~\cite{Raggi:2015yfk,Bauer:2018onh}.%
\footnote{See also Refs.~\cite{Redondo:2008ec,Fradette:2014sza,Kamada:2015era,Kamada:2018zxi,Chang:2016ntp,Escudero:2019gzq,Ibe:2019gpv,Escudero:2020dfa} for cosmological
 and astrophysical constraints. }

In this paper, we discuss
how the topological defects 
in the dark sector affect the SM sector
through the gauge kinetic mixing (see Refs.~\cite{Brummer:2009cs,Long:2014mxa,Hook:2017vyc}
for earlier works).%
\footnote{In this paper, we assume CP-conserving kinetic mixing.
The CP-violating mixing has been discussed in Refs.~\cite{Terning:2018lsv,Terning:2019bhg}.}
In particular, we discuss 
the effects of the topological defects 
in a model where the U(1) gauge symmetry associated with the dark photon 
is embedded in an SU(2) gauge symmetry.
Hereafter, we call them
the $\mathrm{U}(1)_\mathrm{D}$ symmetry and the $\mathrm{SU}(2)_\mathrm{D}$ symmetry, respectively.
The $\mathrm{SU}(2)_\mathrm{D}$ symmetry is spontaneously broken down to the $\mathrm{U}(1)_\mathrm{D}$ symmetry by
a vacuum expectation value (VEV) of a scalar field in the adjoint representation
of SU(2)$_\mathrm{D}$ at a high energy scale.
The $\mathrm{U}(1)_\mathrm{D}$ symmetry is subsequently broken down to ${\mathbb Z}_2$ symmetry by another scalar field 
of the adjoint representation of SU(2)$_\mathrm{D}$ at a lower energy scale.
Here, the ${\mathbb Z}_2$ symmetry is the center of $\mathrm{SU}(2)_\mathrm{D}$.
The dark photon which corresponds to the $\mathrm{U}(1)_\mathrm{D}$ gauge boson obtains a mass at the second symmetry breaking.

When the two scales of the symmetry breaking are hierarchically separated,
we can discuss the phase transitions separately.
At the first symmetry breaking,
there appear the ’t~Hooft-Polyakov magnetic monopoles whose topological charges are the elements of the homotopy group,
$\pi_2\left(\mathrm{SU(2)}_{\mathrm{D}}/\mathrm{U(1)}_\mathrm{D}\right)= {\mathbb Z}$~\cite{tHooft:1974kcl, Polyakov:1974ek, Georgi:1972cj}.
At the second symmetry breaking, the cosmic strings are formed due to $\pi_1\left(\mathrm{U(1)}_\mathrm{D}\right) = {\mathbb Z}$~\cite{Abrikosov:1956sx, Nielsen:1973cs}.
As a notable feature of the successive breaking, $\mathrm{SU(2)}_{\mathrm{D}}\to\mathrm{U(1)}_\mathrm{D}\to\mathbb{Z}_2$, 
the dark magnetic flux of the monopole formed at the first phase transition is confined into the dark cosmic strings at the second phase transition.
Such a composite topological defect 
is called the bead solution~\cite{Hindmarsh:1985xc,Everett:1986eh,Aryal:1987sn,Kibble:2015twa}.
The network of the connected bead solutions is also called the necklace~\cite{Berezinsky:1997td}.%
\footnote{The appearance of the necklace in SO$(10)$ or E$_6$ broken into the SM gauge group is investigated in $e.g.$ Ref.~\cite{Lazarides:2019xai}.}

As discussed in Ref.~\cite{Brummer:2009cs}, the dark ’t~Hooft-Polyakov magnetic monopole does not induce the magnetic field of the QED even in the presence of the kinetic mixing in the $\mathrm{U}(1)_\mathrm{D}$
symmetric phase.
As we will see, however, the dark bead solution leads to the hedgehog shaped QED magnetic flux around the bead, which looks like a QED magnetic monopole. At the same time, the Bianchi identity of QED is satisfied everywhere. We call such a solution, a pseudo magnetic monopole. The solution we construct provides an ultraviolet completion of the system of the coexisting dark monopoles and the dark strings discussed in e.g., Ref.~\cite{Hook:2017vyc}.
We also perform a 3+1 dimensional numerical simulation to see how such a topological defect is formed at the phase transition.

The organization of the paper is as follows. 
In Sec.~\ref{sec:simpledefects}, we discuss the effects of 
the cosmic string and the monopole solutions in the dark sector to the SM sector through the kinetic mixing.
In Sec.~\ref{sec:beads}, we discuss how the dark bead/necklace solutions look in the presence of the kinetic mixing. There, we show that the bead on the string leads to the pseudo magnetic monopole solution of the QED.  
In Sec.~\ref{sec:simulation}, we perform a numerical simulation to see
the formation of the pseudo magnetic monopole.
The final section is devoted to the conclusion.

%--------------------------------------------------------
\section{Gauge Kinetic Mixing and Strings/Monopoles}
\label{sec:simpledefects}
In this section, we discuss how the SM sector is affected 
by the dark cosmic string and the dark  monopole through the kinetic mixing.
In the following, we focus on the effects on the QED gauge field,
although the following discussion can be 
extended to the full SM.
The QED gauge field configuration in the presence of the bead solution is discussed in the next section.
\subsection{Cosmic String}
\label{sec:strings}
First, let us consider the dark photon model based on the $\mathrm{U}(1)_\mathrm{D}$ gauge theory coupling to the QED photon,
\begin{align}
\label{eq:U1model}
\mathcal{L}=-\frac{1}{4}F_{\mu\nu}F^{\mu\nu}-\frac{1}{4}F'_{\mu\nu}F^{'\mu\nu}+\frac{\epsilon}{2}F_{\mu\nu}F^{'\mu\nu}+D_{\mu}\phi D^{\mu}\phi^*-V(\phi)\ .
\end{align}
Here, $F_{\mu\nu}$ and $F'_{\mu\nu}$ represent the gauge field strengths of the $\mathrm{U}(1)_{\mathrm{QED}}$ and 
the $\mathrm{U}(1)_\mathrm{D}$ gauge theories,
respectively.
The corresponding gauge fields 
are given by $A_\mu$ and $A_\mu'$.
The parameter $\epsilon$ is the kinetic mixing parameter.%
\footnote{The kinetic mixing parameter to $\mathrm{U}(1)_{\mathrm{Y}}$ of the SM, i.e., $\epsilon_{{Y}}F^{{Y}}_{\mu\nu}F'^{\mu\nu}/2$, is
related to $\epsilon$ by $\epsilon=\epsilon_Y \cos\theta_{W}$ with $\theta_W$ being the weak mixing angle.}
The gauge coupling constants of $\mathrm{U}(1)_{\mathrm{QED}}$ and $\mathrm{U}(1)_\mathrm{D}$
are denoted by $e$ and $g$, respectively.
Throughout this paper, we assume that the charge assignment of the $\mathrm{U}(1)_{\mathrm{QED}}$ and the $\mathrm{U}(1)_\mathrm{D}$ is exclusive in the basis defined in Eq.\,\eqref{eq:U1model}.
That is, 
all the SM fields are neutral under 
$\mathrm{U}(1)_\mathrm{D}$,
while all the $\mathrm{U}(1)_\mathrm{D}$ charged fields 
are neutral under $\mathrm{U}(1)_{\mathrm{QED}}$.

To break the $\mathrm{U}(1)_\mathrm{D}$, we introduce a complex scalar field $\phi$ with the $\mathrm{U}(1)_\mathrm{D}$ charge $1$.
The covariant derivative of $\phi$ is given by
\begin{align}
D_{\mu}\phi=(\partial_{\mu}-igA'_{\mu})\phi\ . 
\end{align}
The scalar potential of $\phi$ is given by
\begin{align}
V=\frac{\lambda}{4}(|\phi^2|-v^2)^2\ ,
\end{align}
where $\lambda > 0$ is a coupling constant and $v$ is a dimensionful parameter.
At the vacuum, 
$\phi$ obtains a VEV, $\langle \phi \rangle = v$, with which the $\mathrm{U}(1)_\mathrm{D}$
is spontaneously broken.

\subsubsection{String solution for $\epsilon = 0$}
Let us begin with the cosmic string in the absence of the kinetic mixing, i.e.,
$\epsilon=0$.
In this paper, we always take the 
temporal gauge when we discuss static gauge field configurations. 
The static string solution along the $z$-axis is given by the form (see e.g., Ref.~\cite{Vilenkin:2000jqa}),
\begin{align}
\label{eq:ansatz1}
&\phi =v h(\rho)e^{i n\varphi}\ , \\
\label{eq:ansatz2}
&A'_i=-\frac{n}{g}\frac{\epsilon_{ij}x^j}{\rho^2}f(\rho)\ ,~~~~(i,j=1,2)\ ,
\end{align}
where the Cartesian coordinate, $(x^1,x^2,x^3)=(x,y,z)$, is 
related to the cylindrical coordinate via $\varphi={\arctan}(y/x)$ and $\rho=\sqrt{x^2+y^2}$.
The anti-symmetric tensor in the two-dimensional space transverse to the $z$-axis, $\epsilon_{ij}$, is defined by $\epsilon_{12}=1$.%
\footnote{With $d\varphi = - dx^i \epsilon_{ij}x^j/\rho^2$, we may rewrite
Eq.\,\eqref{eq:ansatz2} by $A'_i dx^i = n/g \times f(\rho)d\varphi$.} 
The profile functions, $h(\rho)$ and $f(\rho)$, satisfy the boundary conditions,
\begin{align}
h(\rho)\rightarrow 0\ , ~(\rho\rightarrow0)&\ ,~~~~~h(\rho)\rightarrow 1\ , ~(\rho\rightarrow \infty)\ ,\\
f(\rho)\rightarrow 0\ , ~(\rho\rightarrow 0)&\ ,~~~~~f(\rho)\rightarrow 1\ , 
~(\rho\rightarrow \infty)\ .
\end{align}
The profile functions can be
determined numerically from the field equations of motion,
which approach to $1$ for $\rho \gg (gv)^{-1}$ exponentially.
The winding number, $n \in \pi_1\left(\mathrm{U(1)}_\mathrm{D}\right)$, 
is related to the dark magnetic flux inside the string core,
\begin{align}
\label{eq:Wilson0}
\int d^2 xB_z'=\oint_{\rho\rightarrow \infty} A'_i dx^i=\frac{2\pi n}{g}\ ,
\end{align}
where $B_z'= \epsilon_{ij}F'_{ij}/2$.

\subsubsection{String solution for $\epsilon \neq 0$}
In the presence of the kinetic mixing,
the field equations of the QED and the dark photon are given by,
\begin{align}
    &\partial_\mu F^{\mu\nu} - \epsilon \partial_\mu F^{\prime\mu\nu} = e J_\mathrm{QED}^\nu\ , \\
    & \partial_\mu\tilde{F}^{\mu\nu} = 0 \ ,\\
    &\partial_\mu F^{\prime\mu\nu} - \epsilon \partial_\mu F^{\mu\nu} = g J_\mathrm{D}^\nu\ , \\
    & \partial_\mu \tilde{F}^{\prime\mu\nu} = 0 \ .
\end{align}
Here, $J_\mathrm{QED}^\mu$ and $J_\mathrm{D}^\mu$ denote the charge currents coupling 
to the QED and the dark photons, respectively.
The second and the fourth equations are the Bianchi identities for $\tilde{F}^{(\prime)}_{\mu\nu} = \epsilon_{\mu\nu\rho\sigma}F^{(\prime)\rho\sigma}/2$.

To discuss the vacuum configuration, let 
us take $J_\mathrm{QED}^\mu = 0$.
Around the dark cosmic string, the dark charged current is given by,
\begin{align}
    J_\mathrm{D}^i= i \phi D_i\phi^\dagger - i \phi^\dagger D_i \phi = 2v^2 n \frac{\epsilon_{ij}x_j}{\rho^2}h^2(f-1)\ ,
\end{align}
which is the circular current around the cosmic string.
For $J_\mathrm{QED}^\mu = 0$, we find that the QED gauge field follows the dark photon configuration,
\begin{align}
\label{eq:FinString}
    F^{\mu\nu} = \epsilon F^{\prime \mu\nu}\ ,
\end{align}
with which the dark photon field equation is reduced to 
\begin{align}
\label{eq:darkEQ}
    (1-\epsilon^2)\partial_\mu F^{\prime\mu\nu}  = g J_\mathrm{D}^\nu\ .
\end{align}
The cosmic string solution satisfying Eq.\,\eqref{eq:darkEQ} is identical to 
Eqs.\,\eqref{eq:ansatz1} and \eqref{eq:ansatz2} with rescaled $g$ and $A_\mu'$ by,
\begin{align}
    & g_s  = \frac{g}{\sqrt{1-\epsilon^2}} \ ,\\
    &gA'_{\mu} = g_s A'_{s\mu}\ .
\end{align}
The resultant dark magnetic flux 
for the dark cosmic string with the winding number
$n$ is given by,
\begin{align}
     \oint A'_{s\mu} d x^\mu = \frac{2\pi n }{g_s} \ .
\end{align}

As a result, we find that the dark cosmic string induces the QED magnetic flux along the cosmic string~\cite{Alford:1988sj, Vachaspati:2009jx} and the Wilson loop of QED around the string is given by,
\begin{align}
W_{\mathrm{QED}} = \oint e A_\mu d x^\mu
= \epsilon e \oint  A'_\mu d x^\mu
= \frac{g_s\epsilon e}{g} \oint A'_{s\mu} d x^\mu 
= \frac{2\pi n \epsilon e}{g} \ .
\end{align}
The corresponding Aharonov-Bohm (AB) phase of a particle with the QED charge $q$ is given by $q W_{\mathrm{QED}}$.
Thus, the dark local string becomes the AB string~\cite{Alford:1988sj} through the kinetic mixing.%
\footnote{The irrational AB phase per $2\pi$ is due to the irrationalities of the kinetic mixing and the ratio $e/g$, which is consistent with the compactness of  U(1)$_\mathrm{D}\times$U(1)$_\mathrm{QED}$. }

Note that the string solution satisfying Eq.\,\eqref{eq:FinString}
can be obtained more easily in the canonically normalized 
basis $(X_\mu,X'_\mu)$ defined by,
\begin{align}
\label{eq:shifted}
    &A_\mu = X_\mu + \epsilon A'_\mu \ , \\
    &A_\mu' =\frac{1}{\sqrt{1-\epsilon^2}} X'_\mu\ .
\end{align}
In the canonically normalized basis, there is no kinetic mixing between $X_\mu$ and $X'_\mu$, and hence,
the configuration 
of the shifted QED is trivial around the 
dark string solution of $X'_\mu$, 
i.e., $X_\mu = 0$. 
The magnetic flux of the dark cosmic string is then given by,
\begin{align}
     \oint X'_{s\mu} d x^\mu = \frac{2\pi n }{g_s} \ .
\end{align}

In this shifted basis, 
the AB phases on the QED charged particles
appear through the direct coupling to the dark photon, $X_\mu'$,
\begin{align}
   \mathcal{L} = -\frac{\epsilon e}{\sqrt{1-\epsilon^2}} X'_\mu J^\mu_\mathrm{QED}\ .
\end{align}
Thus, again, the AB phase of the QED charged particle
with the charge $q$ is given by,
\begin{align}
qW_{\mathrm{QED}} = \frac{q\epsilon e}{\sqrt{1-\epsilon^2}}\oint X'_\mu d x^\mu= \frac{2\pi n q \epsilon e}{g} \ .
\end{align}

\subsection{Monopole}
\label{sec:monopoles}
Next, we discuss the effects of
the kinetic mixing
on the 't Hooft-Polyakov-type monopole~\cite{tHooft:1974kcl, Polyakov:1974ek, Georgi:1972cj}.
Unlike in the previous subsection, we here assume that the $\mathrm{U}(1)_\mathrm{D}$ 
gauge symmetry
stems from
$\mathrm{SU}(2)_\mathrm{D}$
gauge symmetry and 
remains unbroken.

\subsubsection{Monopole for $\epsilon=0$}
Let us first consider the $\mathrm{SU}(2)_\mathrm{D}$ gauge theory with a scalar field in the adjoint representation, $\phi^a~(a=1,2,3)$.
The relevant Lagrangian density is given by,
\begin{align}
&\mathcal{L}=-\frac{1}{4}F^{\prime a}_{\mu\nu}F^{\prime a\mu\nu}+\frac{1}{2}D_{\mu}\phi^aD^{\mu}\phi^a-\frac{\lambda}{4}(\phi^a\phi^a-v^2)^2\ , \\
&F^{\prime a}_{\mu\nu}=\partial_{\mu}A_{\nu}^{\prime a}-\partial_{\nu}A_{\mu}^{\prime a}+g\epsilon^{abc}A^{\prime b}_{\mu}A^{\prime c}_{\nu}\ , \\
&D_{\mu}\phi^a=\partial_{\mu}\phi^a+g\epsilon^{abc} A^{\prime b}_{\mu}\phi^c\ .
\end{align}
Here, $A_{\mu}^{\prime a}$ is the $\mathrm{SU}(2)_\mathrm{D}$ gauge field, 
$F^{\prime a}_{\mu\nu}$ its field strength, and $g$ is the gauge coupling constant of $\mathrm{SU}(2)_\mathrm{D}$. 
We assume $\lambda \sim g$ in the following analysis.
As in the previous subsection, 
we also assume that there is no particles
which are charged under both the $\mathrm{SU}(2)_\mathrm{D}$ and the SM gauge group.
At the vacuum, $\mathrm{SU}(2)_\mathrm{D}$ is spontaneously broken down to $\mathrm{U}(1)_\mathrm{D}$
by the VEV of $\phi^a$,
\begin{align}
\label{eq:SU2trivial}
    \langle \phi^a \rangle = v \delta^{a3}\ ,
\end{align}
where the direction of the vector $\langle\phi^a\rangle$ in the $\mathrm{SU}(2)_\mathrm{D}$ space can be chosen arbitrary. 
Once we choose the vacuum in Eq.\,\eqref{eq:SU2trivial}, the gauge potential of the remaining U(1)$_\mathrm{D}$ gauge symmetry corresponds to $A_\mu^3$.

At the phase transition, 
$\mathrm{SU}(2)_\mathrm{D}\rightarrow\mathrm{U}(1)_\mathrm{D}$, the dark monopole appears.
The static monopole solution is given by,
\begin{align}
\label{eq:heg}
&\phi^a=vH(r)\displaystyle{\frac{x^a}{r}}\ , \\
\label{eq:heg2}
&A^{\prime a}_i=
\frac{1}{g}\displaystyle{\frac{\epsilon^{aij}x^j}{r^2}}F(r)\ ,\quad (i,j=1,2,3)\ ,
\end{align}
where $r=\sqrt{x^2+y^2+z^2}$, and $ \epsilon^{aij}$ is the anti-symmetric tensor in the three-dimensional space
with a convention $\epsilon^{123}=1$.
The profile functions, $H(r)$ and $F(r)$ satisfy,
\begin{align}
H(r)\rightarrow 0\ , ~(r\rightarrow0)\ ,~~~~~H(r)\rightarrow 1\ ,~(r\rightarrow \infty)\ ,\\
F(r)\rightarrow 0 \ ,~(r\rightarrow 0)\ ,~~~~~F(r)\rightarrow 1\ ,~(r\rightarrow \infty)\ .
\end{align}
The profile functions can be determined numerically by solving the field equations of motion,
which 
converge exponentially to
the asymptotic values at $r \gg (gv)^{-1}$.

To see how the dark magnetic field emerges, it 
is convenient to define an effective U(1)$_\mathrm{D}$
field strength,
\begin{align}
\label{eq:effectiveF}
\mathcal{F}'_{\mu\nu} \equiv \frac{1}{v}\phi^a F^{\prime a}_{\mu\nu}\ ,
\end{align}
(see e.g. Ref.~\cite{Shifman:2012zz}).
The only non-vanishing components of $\mathcal{F}^{\prime\mu\nu}$ are 
\begin{align}
    \mathcal{F}^{\prime i j } = -
    \frac{1}{g}
\frac{\epsilon^{ijk}x^k}{r^3} (2F-F^2)H \ , \quad (i,j=1,2,3)\ .
\end{align}
Hence, the dark magnetic charge of the monopole solution is given by,
\begin{align}
\label{eq:Mcharge}
    Q_M'=  \frac{1}{2} \int_{r \to \infty} 
    dS_{ij} \mathcal{F}^{\prime i j } =  -\frac{4\pi}{g} \ ,
\end{align}
where $dS_{ij}$ is the surface element of the two dimensional sphere surrounding the monopole.

The monopole solution 
satisfies the Gauss law at the vacuum, that is,
\begin{align}
\label{eq:Gauss}
   & \partial_\mu \mathcal{F}^{\prime\mu\nu} 
= 0 \ .
\end{align}
On the other hand, it satisfied the Bianchi identity,
\begin{align}
\label{eq:bianchi}
&\partial_\mu \tilde{\mathcal{F}}^{\prime\mu\nu} 
=  0\ ,
\end{align}
only at $r\gg (gv)^{-1}$.%
\footnote{The Gauss law is satisfied since 
$\mathcal{F}'_{ij} = \epsilon_{ijk}x_k/r^3\times(2F-F^2)H$  
with the boundary conditions $F(r) \propto r^2$ and $H(r)\propto r$ for $r \to 0$.
The Bianchi identity is satisfied when $F$ and $H$ are constants.  }
Therefore, the dark photon gauge field $A_\mu'$ cannot be defined globally around the monopole.
The monopole solution in terms of the $\mathrm{SU}(2)_\mathrm{D}$
gauge field is, on the other hand, defined globally.

For a later purpose, we introduce a local 
dark photon gauge field defined at $r \gg (gv)^{-1}$.
Let us cover the region of $r \gg(gv)^{-1}$
by two charts of the polar coordinate in the north and the south hemispheres,
\begin{align}
\label{eq:chart}
    U_{N} = 
    \left\{(r, \theta,\varphi)|0 \le \theta \le {\pi}/{2} + \varepsilon
    \right\}\ ,
    \quad 
    U_{S} = 
    \left\{(r, \theta,\varphi)|{\pi}/{2}-\varepsilon \le \theta \le \pi
    \right\}\ ,
\end{align}
Here, $\theta$ is the 
zenith angle and $\varepsilon$ is a tiny positive parameter.
These two charts overlap at around the equator, $\theta = \pi/2$.
In the polar coordinate, 
the monopole configuration in
Eqs.\,\eqref{eq:heg} and \eqref{eq:heg2} at $r \gg (gv)^{-1}$ are given by,
\begin{align}
&\phi^a  \to v(s_\theta c_\varphi, s_\theta s_\varphi, c_\theta)\ , \\
&A_r^{\prime a} \to 0\ , \\
&A_\theta^{\prime a} \to \frac{1}{g}(s_\varphi,-c_\varphi,0)\ ,\\
&A_\varphi^{\prime a} \to \frac{1}{g}(s_\theta c_\theta c_\varphi,  s_\theta c_\theta s_\varphi, -s_\theta^2)\ ,
\end{align}
where each component in the right-hand side corresponds to $a=1,2,3$.
We abbreviate $\cos$ and $\sin$ by 
$c$ and $s$, respectively.
The gauge fields in the polar coordinates,
$A^{\prime a}_{r, \theta,\varphi}$, are read off from
$A^{\prime a}_idx^i = A^{\prime a}_{r}dr + A^{\prime a}_{\theta}d\theta + A^{\prime a}_{\varphi}d\varphi$.
This expression is valid in the both charts.

To see the relation with the monopole configuration with the trivial vacuum in Eq.\,\eqref{eq:SU2trivial},
let us perform a SU(2)$_\mathrm{D}$ gauge transformation in each chart given by,
\begin{align}
\label{eq:NStransform}
    &g_N =\left(
    \begin{array}{cc}
      c_{\theta/2}  &  e^{-i\varphi}s_{\theta/2} \\
      - e^{i\varphi}s_{\theta/2}& c_{\theta/2}
    \end{array}
    \right)\ ,
    ~~~~~~~ g_S =\left(
    \begin{array}{cc}
       e^{i\varphi}c_{\theta/2}  & s_{\theta/2} \\
      -s_{\theta/2}& e^{-i\varphi}c_{\theta/2}
    \end{array}
    \right)\ .
\end{align}
In each chart,  $\phi^a$ and $A^a_i$ are transformed to 
\begin{align}
    &\phi^a\tau^a \to  \phi_{N,S}^{a}\tau^a = g_{N,S} \phi^a\tau^a g_{N,S}^\dagger\ , \\
&A^{\prime a}_i\tau^a \to  A^{\prime a}_{N,S\,i}\tau^a = g_{N,S} A^{\prime a}_i\tau^a g_{N,S}^\dagger - 
\frac{i}{g}(\partial_i g_{N,S}) g_{N,S}^\dagger\ ,
\label{eq:localDP}
\end{align}
where $\tau^{a=1,2,3}$ denote the half of the Pauli matrices.

In the $U_N$ chart, the asymptotic behaviors are given by,
\begin{align}
\label{eq:phiN}
    &\phi^a \to \phi_N^a = v \delta^{a3}\ ,\\
    & A^{\prime a} \to A^{\prime a}_{N} = \frac{1}{g}\delta^{a3} (\cos\theta -1) d\varphi\ ,
\end{align}
with $A^{\prime a}_{N\,r,\theta}$  vanishing.
In the $U_S$ chart, they are given by,
\begin{align}
\label{eq:phiS}
    &\phi^a \to \phi^a_S  = v \delta^{a3}\ ,\\
    & A^{\prime a} \to A^{\prime a}_{S}= \frac{1}{g}\delta^{a3} 
    (\cos\theta+1)  d\varphi\ ,
\end{align}
with $A^{\prime a}_{S\,r,\theta}$ vanishing.
After the gauge transformation, the configurations $\phi_{N,S}^a$ are along the $a=3$ direction as in Eq.\,\eqref{eq:SU2trivial} in both charts.
Accordingly, the dark photon gauge potential corresponds to $A^{\prime 3}_{N,S}$
as in the case of the trivial vacuum.
In the following, we call this gauge choice in the two charts the combed gauge.

In the combed gauge, $A^{\prime a}$ is defined not globally but only locally by $A^3_{N,S}$ in each chart.  They are connected with each other at around the equator $\theta\sim \pi/2$  by
\begin{align}
    A^{\prime 3}_{S}= A^{\prime 3}_{N}+ \frac{2}{g} d\varphi\ .
\end{align}
In other word, the two charts of the U(1) bundle are connected by the U(1) gauge transition function from $U_N$ to $U_S$,
\begin{align}
    \label{eq:transition}
    t_{NS} = e^{2i\varphi}\ ,
\end{align}
at around the equator.
Since the minimal electric charge of U(1)$_\mathrm{D} \subset$ SU(2)$_\mathrm{D}$ 
is $1/2$ in the unit of $g$, this transition function corresponds to the 
magnetic monopole with a minimal magnetic charge, $Q_M = 4\pi/g$, so that  the Dirac quantization condition is satisfied.%
\footnote{The transition function at the equator, $t_{NS}=e^{2in\varphi}$ ($n\in \mathbb{Z}$), corresponds to the magnetic charge $Q_M = 4\pi n/g$. }
The field strength of $A^{\prime 3}_{S,N}$, on the other hand, coincides with 
\begin{align}
    F^{\prime 3} \to -\frac{1}{g} \sin \theta\, d\theta\wedge d\varphi\ ,
\end{align} 
at $r \gg (gv)^{-1}$ in both the charts.

\subsubsection{Monopole for $\epsilon \neq 0$}
\label{sec:MonopoleEpsilon}
In the non-Abelian extension of the dark sector,
the kinetic mixing between  $\mathrm{U}(1)_\mathrm{D} \subset \mathrm{SU}(2)_\mathrm{D}$
and $\mathrm{U}(1)_\mathrm{QED}$ 
originates from a higher dimensional term,
\begin{align}
\label{eq:SU2model}
    \mathcal{L}_\mathrm{gauge} =
    -\frac{1}{4}F_{\mu\nu}F^{\mu\nu}-\frac{1}{4}F^{\prime a}_{\mu\nu}F^{\prime a\mu\nu}+\frac{\phi^a}{2\Lambda} F^{\prime a}_{\mu\nu}F^{\mu\nu}\ ,
\end{align}
where $\Lambda$ is a high-energy cutoff scale at $\Lambda \gg v$.
Here, we show only the kinetic and the kinetic mixing terms of the gauge fields.
Similarly to the case of the $\mathrm{U}(1)_\mathrm{D}$ model, we assume that no fields are charged under both the SM and the
$\mathrm{SU}(2)_\mathrm{D}$ gauge symmetries.

At the trivial vacuum in Eq.\,\eqref{eq:SU2trivial},
the above higher dimensional operator 
provides the kinetic mixing parameter,
\begin{align}
\label{eq:effectivee}
    \epsilon = \frac{v}{\Lambda}\ ,
\end{align} 
where $A^{\prime 3}_\mu$ is identified with the dark photon, $A_\mu'$. 
For $\epsilon \ll 1$, the effect of the kinetic mixing to the scalar configuration is expected to be negligible.
In this case, we may go to the shifted basis $(X_\mu,X'_\mu)$ defined in Eq.\,\eqref{eq:shifted} to cancel the kinetic mixing term.
In the presence of the dark monopole, on the other hand, the kinetic mixing term is
no more constant in the three-dimensional space,
\begin{align}
    \mathcal{L}_{\mathrm{mixing}} =-
\frac{v}{2\Lambda}\frac{x^a}{r}H(r)F^{\prime a}_{\mu\nu}F^{\mu\nu}\ .
\end{align}
%Besides, the dark photon gauge field $A_\mu'$ is not globally defined around the monopole solution.
Thus, the kinetic mixing term cannot be cancelled by shifting the QED gauge boson.%
\footnote{In the combed gauge in Eqs.\,\eqref{eq:phiN} and \eqref{eq:phiS},
the gauge kinetic term is a constant in the asymptotic region, $r \gg (gv)^{-1}$.
In this case, we can define a shifted QED gauge boson in each chart and can cancel the kinetic mixing term, though it is not possible to cancel the mixing term globally.}

Now let us look at the field equations of the QED gauge field around the dark monopole solution;
\begin{align}
&\partial_\mu F^{\mu\nu} - \epsilon \partial_\mu \mathcal{F}^{\prime\mu\nu} = e J_\mathrm{QED}^\nu\ , \\
    & \partial_\mu\tilde{F}^{\mu\nu} = 0 \ .
\end{align}
Here, we have used the definitions in Eqs.\,\eqref{eq:effectiveF} and \eqref{eq:effectivee}.
Then, 
by remembering the Gauss law of $\mathcal{F}^{\prime \mu\nu}$ in Eq.\,\eqref{eq:Gauss},
we find that the QED field satisfies the
equation of motion
without the dark gauge field. Thus, we conclude that no QED field
is induced
around the dark monopole~\cite{Brummer:2009cs}.

In summary,
\begin{itemize}
    \item The dark cosmic string induces the QED gauge flux, 
    $F_{\mu\nu} = \epsilon F'_{\mu\nu}$,
    in the basis defined in Eq.\,\eqref{eq:U1model}.
    The dark cosmic string induces the AB phases of the QED charged particles through the kinetic mixing.
    \item The dark magnetic monopole does not induce the QED gauge flux, i.e. $F_{\mu\nu}=0$, 
    in the basis defined in Eq.\,\eqref{eq:SU2model}.
    The QED charged particles do not interact with the dark monopole.
\end{itemize}

\section{GAUGE KINETIC MIXING AND BEADS}
\label{sec:beads}
In the previous section, we 
discussed the effects of the gauge kinetic mixing around the 
string solution and the monopole solution.
In this section, we discuss 
the effects around the so-called bead solution which appears
at the successive spontaneous breaking,
SU(2)$_\mathrm{D} \to \mathrm{U}(1)_\mathrm{D} \to \mathbb{Z}_2$ (see Ref.~\cite{Kibble:2015twa} for review).
As we will review shortly,
the magnetic monopole formed 
at the first phase transition 
becomes a seed of the cosmic string at the second phase transition.
The dark magnetic flux of the monopole is confined in the cosmic string.
The bead solution is one of such configuration in which a string and an anti-string are attached
to a monopole.

\subsection{Bead solution for $\epsilon = 0$}
\label{sec:review}

We first consider the bead solution
in the absence of the kinetic mixing.
To achieve the successive symmetry breaking of $\mathrm{SU}(2)_\mathrm{D}$, we introduce 
two adjoint representation scalar fields, $\phi_1^a$ and $\phi_2^a$ $(a=1,2,3)$. 
The potential of these scalar fields is assumed to be
\begin{align}
\label{eq:potential}
	V = \frac{\lambda_1}{4}
	(\phi_1\cdot\phi_1-v_1^2) + \frac{\lambda_2}{4}(\phi_2\cdot\phi_2-v_2^2) + \frac{\kappa}{2}(\phi_1\cdot \phi_2)^2 \ ,
\end{align}
where $\phi_i\cdot\phi_j=  \phi_i^a\phi_j^a$.
We omit terms such as $(\phi_1\cdot\phi_1)(\phi_2\cdot\phi_2)$,
for simplicity.%
\footnote{As for the terms such as $(\phi_1\cdot\phi_2)$, $(\phi_1\cdot\phi_1)(\phi_1\cdot \phi_2)$ and 
$(\phi_1\cdot\phi_2)(\phi_2\cdot \phi_2)$ may be suppressed 
by additional $\mathbb{Z}_2$ 
symmetry under which $\phi_1$ is odd while $\phi_2$ is even.
This additional symmetry remains
unbroken  by the VEV of $\phi_1$ in Eq.\,\eqref{eq:VEV1} in combination with the $\mathbb{Z}_2$ element in $\mathrm{SU}(2)_\mathrm{D}/\mathrm{U}(1)_\mathrm{D}$. }
The coupling constants, $\lambda_1, \lambda_2$ and $\kappa$ are taken to be positive.
We assume $v_1 \gg v_2$, 
so that the
intermediate $\mathrm{U}(1)_\mathrm{D}$ symmetric phase is meaningful.

When $\phi_1^a$ takes the trivial vacuum configuration
\begin{align}
\label{eq:VEV1}
    \langle{\phi_1^a}\rangle = v_1 \delta^{a3}\ ,
\end{align}
 $\mathrm{SU}(2)_\mathrm{D}$   is broken down to the $\mathrm{U}(1)_\mathrm{D}$.
The remaining $\mathrm{U}(1)_\mathrm{D}$ symmetry corresponds to the SO(2) rotation around the $a=3$ axis of SO(3)$\simeq\mathrm{SU}(2)_\mathrm{D}$ vectors, $\phi_{1,2}^a$.
Subsequently,
$\phi_2$ obtains a non-vanishing VEV at a much lower energy scale to minimize the second term of the potential.
For $\kappa>0$, 
the last term in Eq.\,\eqref{eq:potential} 
lifts the $a=3$ component of $\phi_2$.
Thus, for $\kappa >0$, the VEV of $\phi_2^a$ is required to be orthogonal to 
$\langle\phi_1^a\rangle$, i.e., $\langle\phi_1\rangle \cdot \langle \phi_2 \rangle = 0$.
As a result, $\langle\phi_2^a\rangle$ takes a value in the $(\phi_2^1,\phi_2^2)$ plane, that is, 
\begin{align}
    \langle\phi_2^a\rangle = v_2 \delta^{a1}\ .
\end{align}
for example,
which breaks $\mathrm{U}(1)_\mathrm{D}$ spontaneously.
In this way, successive symmetry breaking,
$\mathrm{SU}(2)_\mathrm{D}\to \mathrm{U}(1)_\mathrm{D} \to \mathbb{Z}_2$,
is achieved for $v_1 \gg v_2$.
Here, the remaining $\mathbb{Z}_2$ 
symmetry is the center of $\mathrm{SU}(2)_\mathrm{D}$ which leaves 
the VEVs of $\phi_{1,2}$ invariant.

Now, let us assume that
the dark monopole is 
formed at the first
phase transition, SU(2)$_\mathrm{D} \to \mathrm{U}(1)_\mathrm{D}$.
The asymptotic form of $\phi_1^a$ 
in the monopole solution
is given by Eq.\,\eqref{eq:heg}, 
\begin{align}
\label{eq:asymp1}
    \phi_1^a \to v_1 \frac{x^a}{r} \ ,
\end{align}
at $r\gg (gv_1)^{-1}$.
At the second stage of the 
phase transition,
the configuration of $\phi_2^a$ prefers a direction orthogonal to $\phi_1^a$ for $\kappa > 0$,
\begin{align}
\label{eq:asymp2}
    \phi_2\cdot\phi_1 \to  0\ ,
\end{align}
at $r\gg (gv_1)^{-1}$.
Around the hedgehog solution
in Eq.\,\eqref{eq:asymp1},
Eq.\,\eqref{eq:asymp2} requires that $\phi_2^a$ is on the tangent space of the sphere with a 
constant amplitude, $|\phi_2| = v_2$.
Such a configuration of 
$\phi_2$ is, however, impossible due to the
Poincar\'e--Hopf (hairy ball) theorem.
Thus, $|\phi_2|$ cannot be a constant everywhere at $r\to \infty$.

To see what kind of $\phi_2$ configuration
is formed, it is convenient to discuss in 
the combed gauge 
introduced
in the previous section. There, the monopole solution of $\phi_1$ behaves 
\begin{align}
\langle\phi_1^a \rangle \to v_1 \delta^{a3}\ ,
\end{align}
at $r \gg (gv_1)^{-1}$ in each hemisphere.
The $\mathrm{U}(1)_\mathrm{D}$ symmetry corresponds to the rotation
around the third axis of $\mathrm{SU}(2)_\mathrm{D}$ as 
in the case of the trivial vacuum.
In the combed gauge, it is useful to define a complex scalar,
\begin{align}
    \tilde{\phi}  = \frac{1}{\sqrt{2}} \left(\phi_2^1 - i \phi_2^2\right)\ ,
\end{align}
whose 
$\mathrm{U}(1)_\mathrm{D}$ charge is $+1$ (see the Appendix~\ref{sec:cov}).
The third component $\phi_2^3$ is fixed to $\phi_2^3 = 0$ due to the condition of $\phi_1\cdot\phi_2=0$.

First, let us suppose that $\tilde{\phi}$ (i.e., $\phi_2^a$) takes a trivial vacuum configuration 
in the north hemisphere at least for $r \gg (gv_2)^{-1}$,
\begin{align}
\label{eq:phiNtrivial}
    &\tilde{\phi}_N = \frac{v_2}{\sqrt{2}}\ . 
 \end{align}
In this trivial configuration, the U(1)$_\mathrm{D}$ gauge flux is expelled by the Meissner effect.
Thus, we may consider a trivial dark gauge field configuration,
\begin{align}
& A^{\prime 3}_{Ni} = 0\ ,
\end{align}
in the north hemisphere at $r\gg(gv_2)^{-1}$.
In the south hemisphere, this configuration is connected to
\begin{align}
\label{eq:phiS2}
    &\tilde{\phi}_S = e^{2i\varphi} \tilde{\phi}_N = e^{2i\varphi} \frac{v_2}{\sqrt{2}}\ , \\
    & A^{\prime 3}_{Si}dx^i = \frac{2}{g}d\varphi\ ,
\end{align}
for $r\gg (gv_2)^{-1}$ due to the transition function in Eq.\,\eqref{eq:transition} at the equator.
Thus, we find that the trivial configuration in the north hemisphere leads to a 
non-trivial winding of $\phi_S$.
Accordingly, we find that the  dark magnetic flux,
\begin{align}
\label{eq:fluxS}
    \oint  A^{\prime 3}_{Si}dx^i = \frac{4\pi}{g}\ , 
\end{align}
is induced in the south hemisphere.
The non-vanishing dark magnetic flux
in the south hemisphere is expected since the expelled magnetic flux
of the monopole from the north hemisphere has to go somewhere to satisfy the Bianchi identity in Eq.\,\eqref{eq:bianchi} for $r \gg(g v_1)^{-1}$.

%-------------------
%------
\begin{figure}[t]
\centering
\hspace{1cm}
\includegraphics[width=5.7cm,pagebox=cropbox,clip]{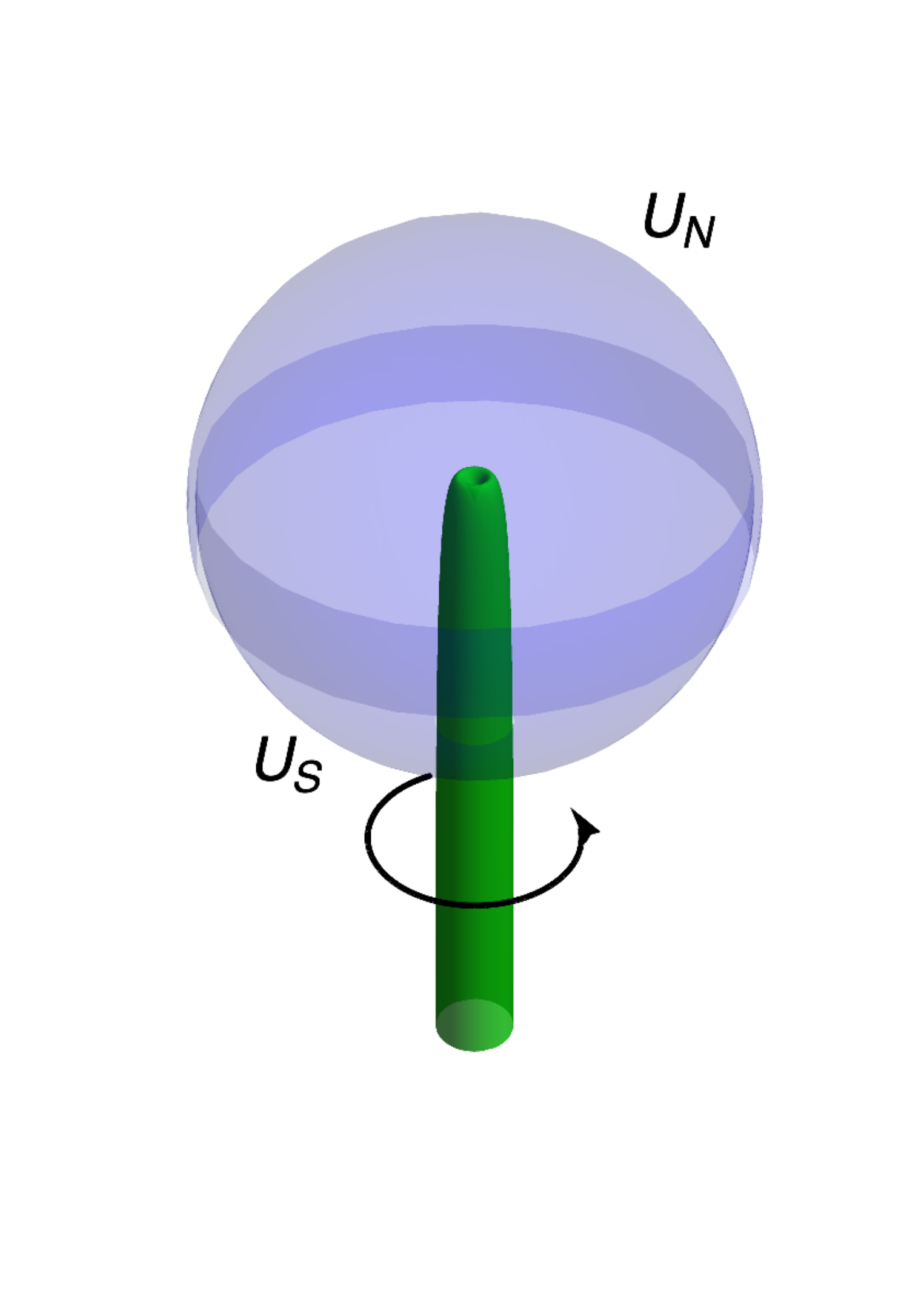}
\hspace{1cm}
\mbox{\raisebox{10mm}{\includegraphics[width=7.5cm,clip]{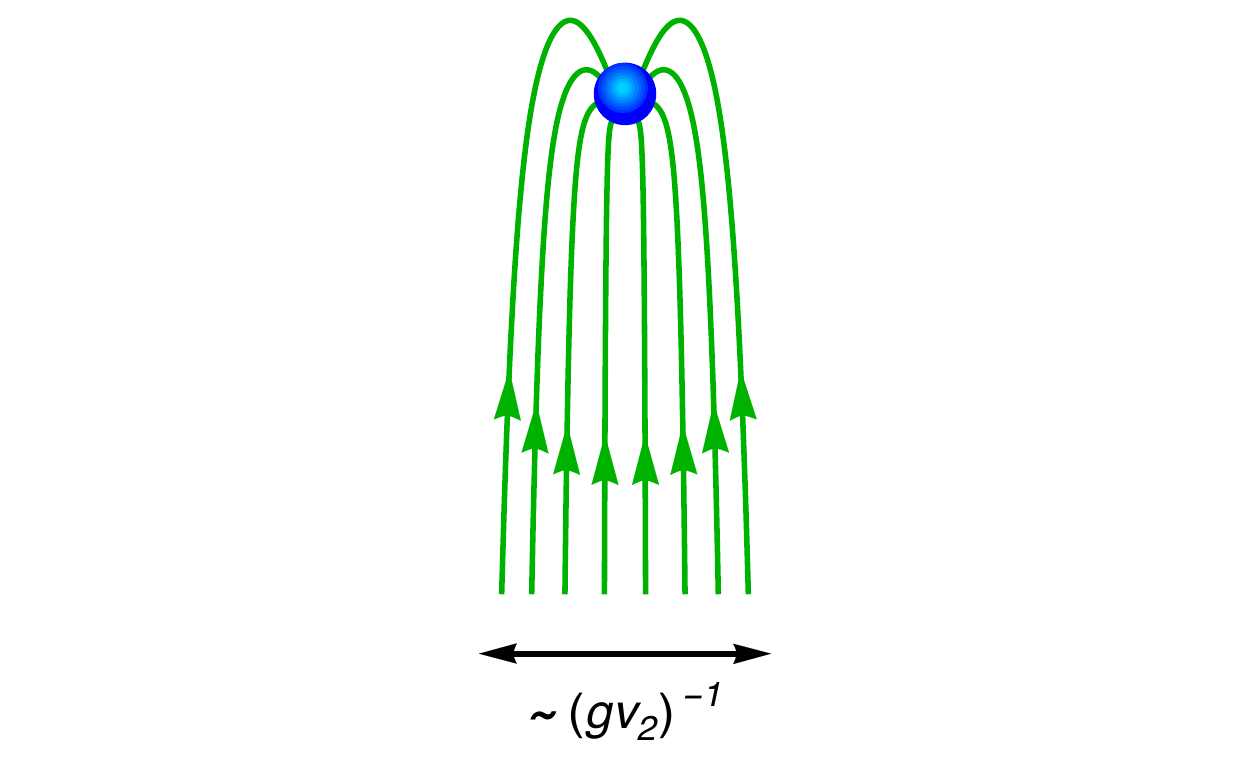}}}
\caption{\sl \small Left) A schematic picture of the string configuration attached to a monopole. The sphere is shown 
as an eye-guide and is divided into the two hemispheres. 
The monopole is placed at the center of the sphere. The attached cosmic string  with $n=2$
extends to $z<0$ direction (green rod).
The arrow shows the direction of the circular current inside the string.
Right) The magnetic flux of
$\mathrm{U}(1)_\mathrm{D}$ in the configuration. 
The magnetic flux is spherical for $r \gg (gv_2)^{-1}$ while it is confined into the cosmic string. The size of the magnetic monopole (the blue point) is of $\sim{(gv_1)^{-1}}$. 
Inside the monopole, the magnetic flux of $\mathrm{U}(1)_\mathrm{D}$ is not well-defined. }
\label{fig:string2}
\end{figure}

%-------------------

In the Higgs phase, the minimum energy solution which carries the magnetic flux is given by the string solution.
In fact, the asymptotic behavior of $\phi_S$ in Eq.\,\eqref{eq:phiS2} coincides with 
that of the cosmic string along the $z$-axis with $n=2$ (see Eq.\,\eqref{eq:ansatz1}).
Therefore, we conclude that the magnetic flux in Eq.\,\eqref{eq:fluxS} is confined in the  cosmic string solution with $n=2$ in the south hemisphere.
The singular asymptotic behavior of $\phi_S$ at $\theta = \pi$ in Eq.\,\eqref{eq:phiS2} is resolved 
by the profile function $h(\rho)$ 
of the cosmic string (see Eq.\,\eqref{eq:ansatz1}).
In Fig.\,\ref{fig:string2}, we show a schematic picture of the magnetic monopole attached by a cosmic string 
with $n=2$.
In the figure, the magnetic flux is spherical for $r \ll (gv_2)^{-1}$, while it is
confined into a cylindrical cosmic string along the $z$ direction extended 
to the $z<0$ region.
Note that this configuration is not static, and the monopole at the center is pulled by the string tension towards the $z<0$ direction.

Another interesting possibility is to suppose that
$\phi$ takes a cosmic string configuration with $n = -1$
in the north hemisphere ($r \gg (gv_2)^{-1}$). 
In this case, the asymptotic configuration of $\tilde\phi$ in the north hemisphere is given by
\begin{align}
\label{eq:phiNstring}
    \tilde{\phi}_N = \frac{v_2}{\sqrt{2}}e^{-i\varphi} h(\rho)\ ,
\end{align}
which is accompanied by
\begin{align}
\label{eq:phiNstring2}
    A^{\prime 3}_{Ni}dx^i  \to -\frac{1}{g} d\varphi\ .
\end{align}
This configuration is connected to 
\begin{align}
\label{eq:phiSstring}
    &\tilde\phi_S \to e^{2\varphi}\tilde\phi_N =\frac{v_2}{\sqrt{2}}e^{i\varphi}\ , \\
\label{eq:phiSstring2}
    &  A^{\prime 3}_{S i}dx^i  \to A^{\prime 3}_{N i}dx^i + \frac{2}{g} d\varphi =  \frac{1}{g} d\varphi\ ,
\end{align}
in the south hemisphere.
The connected configuration is nothing but the cosmic string solution with $n=1$.
Thus, totally, this configuration has a string and an anti-string configurations attached 
to a magnetic monopole.
The magnetic fluxes  confined in the string and the anti-string 
sum up to
\begin{align}
\label{eq:beadflux}
  \oint A^{\prime 3}_{Ni}dx^i -\oint A^{\prime 3}_{Si}dx^i =  - \frac{4\pi}{g} \ ,
\end{align}
which coincides with the magnetic flux of the monopole in Eq.\,\eqref{eq:Mcharge}.

%-------------------
%------
\begin{figure}[t]
\centering
\hspace{-2.2cm}
\includegraphics[width=5.7cm,pagebox=cropbox,clip]{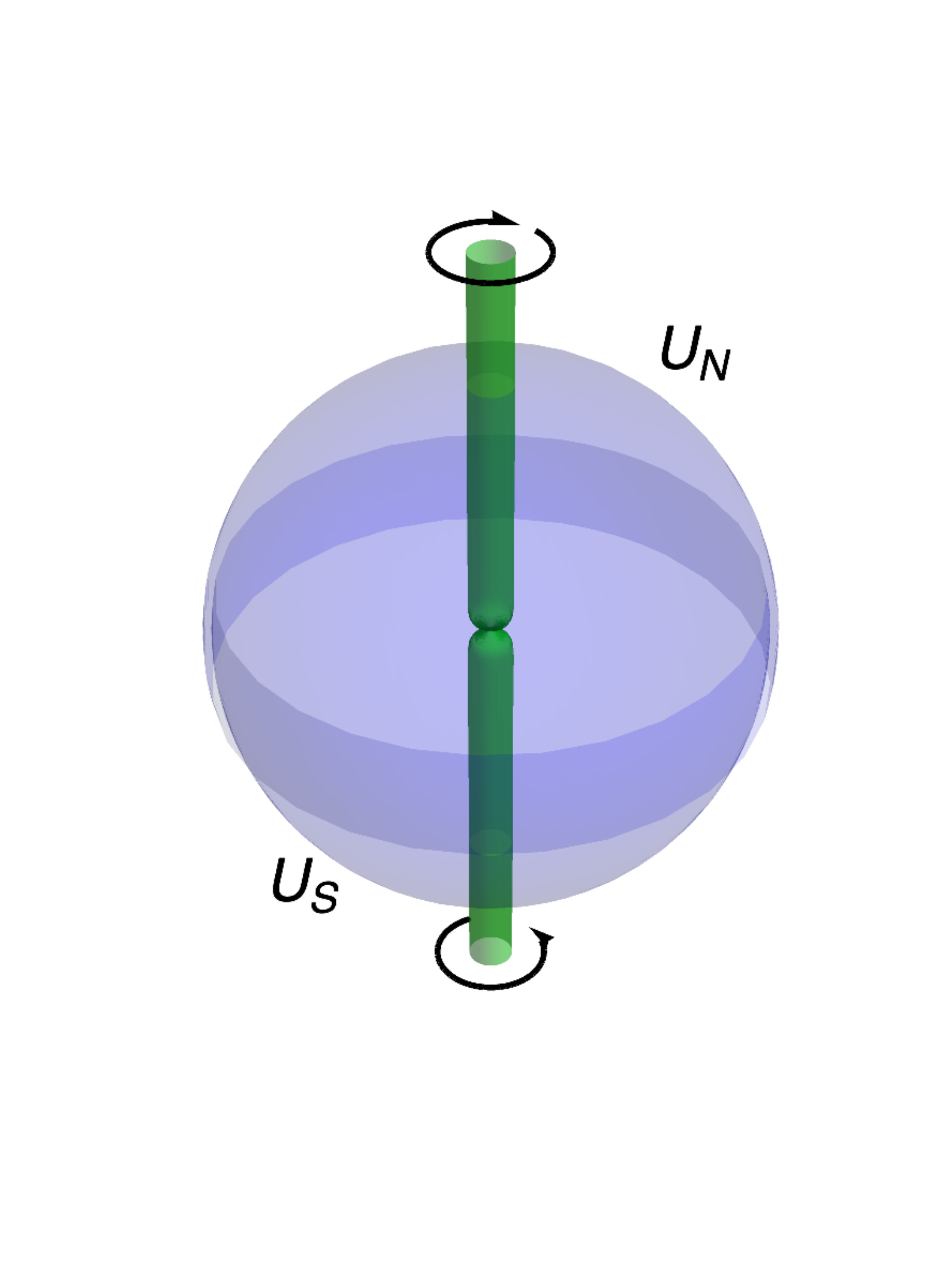}
\hspace{4.2cm}
\mbox{\raisebox{0mm}{\includegraphics[width=1.4cm,clip]{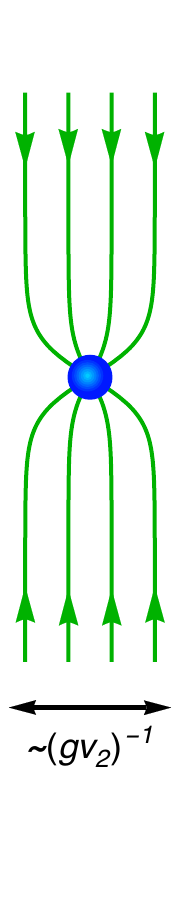}}}
\caption{\sl \small Left) A schematic figure of the bead configuration. 
A string and an anti-string are
attached to the 
monopole placed at the center of the sphere.
The arrows show the directions of the circular currents inside the strings.
Right) The magnetic flux of
$\mathrm{U}(1)_\mathrm{D}$ in the configuration. }
\label{fig:bead}
\end{figure}

The above configuration is called the bead solution~\cite{Hindmarsh:1985xc}.
In Fig.\,\ref{fig:bead}, we show a schematic picture of the 
bead solution.
As in the case of Fig.\,\ref{fig:string2}, the magnetic flux is
expected to be 
spherical for 
$r \ll (gv_2)^{-1}$, while it is
confined into cylindrical cosmic strings along the $z$ direction.
 The bead solution
is static unlike in the case of the monopole attached by one string with $n=2$, since the monopole is pulled by the same force by the two strings into 
opposite directions.
The network of the bead solution is also called the necklace.

So far, no analytic expression of the bead solution has been known.
Instead, we refer  three-dimensional classical lattice simulations of the SU(2) gauge field theory with two adjoint scalar fields. 
Those works have confirmed formation of the beads and the necklaces \cite{Hindmarsh:2016dha} (See also Ref.~\cite{Siemens:2000ty} for other numerical simulation of the necklaces). The cosmological evolution of necklaces have also been explored in Refs.~\cite{Berezinsky:1997td, Siemens:2000ty, BlancoPillado:2007zr, Hindmarsh:2016dha}.
We also show our results of the numerical simulation in the next section.

In the above discussion, 
we have used the locally defined gauge fields
in the north and the south hemispheres.
It is informative to go to the 
gauge where
the $\mathrm{SU}(2)_\mathrm{D}$ gauge fields are globally defined.
In fact, the globally defined $\mathrm{SU}(2)_\mathrm{D}$ fields are obtained by undoing $g_N$ transformation,
\begin{align}
    &\phi_2^a\tau^a = g_{N}^\dagger \phi_{2N}^a\tau^a g_{N}\ ,\\
    &A_i^{\prime a}\tau^a = g_{N}^\dagger A_{Ni}^{\prime a}\tau^a g_{N}-
    \frac{i}{g}
    (\partial_i g_N^\dagger)g_{N}
\ ,
\end{align}
in the north hemisphere and
by undoing $g_S$ transformation,
\begin{align}
    &\phi_2^a\tau^a = g_{S}^\dagger \phi_{2S}^a\tau^a g_{S}\ ,\\
    &A_i^{\prime a}\tau^a = g_{S}^\dagger A_{Si}^{\prime a}\tau^a g_{S}-
    \frac{i}{g}
    (\partial_i g_S^\dagger)g_{S}
\ ,
\end{align}
in the south hemisphere.
These fields coincide at the equator, $\theta = \pi/2$, and hence, they are connected smoothly.
Note that 
they become equal not only at the equator but also for $0\le \theta\le \pi$ in the asymptotic region, $r \gg (gv_2)^{-1}$ and 
$\rho \gg(gv_2)^{-1}$.
For example, the asymptotic behaviors of the bead solution are given by,
\begin{align}
\label{eq:asympphi1}
&\phi_1^a  \to v_1(s_\theta c_\varphi, s_\theta s_\varphi, c_\theta)\ , \\
\label{eq:asympphi2}
&\phi_2^a  \to v_2(c_\theta c_\varphi, c_\theta s_\varphi, -s_\theta)\ , \\
&A_r^{\prime a} \to 0\ , \\
&A_\theta^{\prime a} \to \frac{1}{g}(s_\varphi,-c_\varphi,0)\ ,\\
&A_\varphi^{\prime a} \to \frac{1}{g}(0,  0, -1)\ .
\end{align}
These asymptotic fields of course satisfies
$D_\mu\phi_1 \to 0$, $D_\mu \phi_2 \to 0$, and $\phi_1\cdot \phi_2 \to 0$ for $r\gg (gv_2)^{-1}$ and $\rho \gg (gv_2)^{-1}$.

Before closing this subsection, let us comment on the topological property of the bead and string solutions.
For the successive symmetry breaking,
$\mathrm{SU}(2)_\mathrm{D} \to \mathrm{U}(1)_\mathrm{D} \to \mathbb{Z}_2$,
the topological property of the vacuum configuration is classified by
 $\pi_1\!\left(\mathrm{SU}(2)_\mathrm{D}/\mathbb{Z}_2\right)=\mathbb{Z}_2$.
The topological defects associated with  $\pi_1\!\left(\mathrm{SU}(2)_\mathrm{D}/\mathbb{Z}_2\right)$ are
cosmic strings with the winding number
\begin{align}
    n = 0, 1\ (\mbox{mod 2})\ .
\end{align}
Other cosmic strings with even and odd winding numbers are topologically
equivalent to the solution with $n=0$ and $n=1$, respectively.
Thus, for example, there should be a continuous path 
which connects the string ($n=1$) and the anti-string ($n=-1$).
The bead solution is the realization of such a path in the three dimensional space, where 
the cosmic string with $n=1$ beneath the 
monopole is flipped to that of $n=-1$ above the monopole.
From this property, the bead solution is also an example of  the junction of the 
$\mathbb{Z}_k$-string~\cite{Tong:2003pz, Shifman:2004dr}.

The configuration with a monopole attached by a cosmic string with the winding number $n=2$ is 
also important for 
the equivalence between 
the trivial vacuum and 
the cosmic strings with even winding
numbers.
In fact, this configuration allows a long string with $n=2$ to break up by creating
a pair of a monopole and an anti-monopole by quantum tunneling~\cite{Preskill:1992ck}.

Note also that the topological charge, $\pi_2(\mathrm{SU}(2)_\mathrm{D}/\mathrm{U}(1)_\mathrm{D})$, is effectively conserved for the successive symmetry breaking,
$\mathrm{SU}(2)_\mathrm{D} \to \mathrm{U}(1)_\mathrm{D} \to \mathbb{Z}_2$.
That is, when the magnetic monopole is formed at the first phase transition, the total magnetic flux measured at the infinite sphere is not changed even if U$(1)_\mathrm{D}$ is spontaneously broken.
The effective conservation of the topological charge, however, does not necessarily lead to the lowest energy configuration (stable vacuum configuration) of a non-trivial element of $\pi_2(\mathrm{SU}(2)_\mathrm{D}/\mathrm{U}(1)_\mathrm{D})$.
For example, let us consider the monopole with the winding number $n=2$. 
If the monopole is attached by two cosmic strings with the winding number $n=\pm 2$ in the opposite direction, the pair creation of the monopole-anti-monopole on the attached cosmic string ends up with the disappearance of the monopole solution with $n=2$.
If, on the other hand, the monopole is attached by four cosmic strings with the winding number $n=\pm 1$, the pair creation does not occur, and hence, the configuration is stable.
Either of the above two configurations will be realized, depending on the interaction between the cosmic strings, i.e., repulsive or attractive.
The interaction between the strings in turn depends on the model parameters~\cite{Bettencourt:1994kf}.

Finally, we comment on the case where $\mathrm{U}(1)_\mathrm{D}$ is broken not by $\phi_2$ but by a Higgs field in the fundamental representation of $\mathrm{SU}(2)_\mathrm{D}$, $\mathrm{SU}(2)_\mathrm{D}$ is completely broken leaving no $\mathbb{Z}_2$ symmetry. 
In this case, the magnetic flux of the cosmic string with $n=1$ is $4\pi/g$.
Therefore, the cosmic string with the minimum winding number, $n=1$, can be broken up by creating a pair of the monopole-anti-monopole pair.

\subsection{Bead Solution for $\epsilon \neq 0$}
\label{sec:kinetic}
As we have seen in the previous section, the
dark cosmic string induces 
the QED magnetic flux inside the string
through the kinetic mixing (see Eq.\,\eqref{eq:FinString}).
On the other hand, the dark monopole
in the $\mathrm{U}(1)_\mathrm{D}$ symmetric 
phase does not induce the QED magnetic flux even in the presence of the kinetic mixing.
In this subsection, we discuss how the dark bead solution affects the QED flux through the kinetic term.

Let us consider a bead solution along the $z$-axis
with the monopole at the origin.
At $z \gg (g v_2)^{-1}$, 
the configuration is the anti-string
given by Eqs.\,\eqref{eq:phiNstring} and \eqref{eq:phiNstring2}.
Thus, in this region,
the dark anti-string induces the 
QED magnetic flux,
\begin{align}
    F_N = \epsilon F^{\prime 3}_{N}= - \frac{\epsilon}{g} \frac{df(\rho)}{d\rho} d\rho\wedge d\varphi \ ,
\end{align}
at $z \gg (g v_2)^{-1}$ (see Eq.\,\eqref{eq:FinString}).
Similarly, the dark string induces the QED magnetic flux,
\begin{align}
    F_S = \epsilon F^{\prime 3}_{S} =  \frac{\epsilon}{g} \frac{df(\rho)}{d\rho} d\rho\wedge d\varphi \ .
\end{align}
at $z \ll-(gv_2)^{-1}$.
Therefore, we find that the QED magnetic flux along the string (in the north hemisphere of the dark monopole) and the 
anti-string (in the south hemisphere) flows into the region $r < \order{(g v_2)^{-1}}$,
which amounts to%
\footnote{The sign of the integration of the surface integration 
is flipped from the one in Eq.\,\eqref{eq:beadflux} since we are interested in the flux flowing into $z \sim (gv_2)^{-1}$.
}
\begin{align}
\int F\big|_{|z| \gg (g v_2)^{-1}} = -\epsilon\oint A^{3}_{Ni}dx^i + \epsilon\oint A^{\prime 3}_{Si}dx^i =  \frac{4\pi \epsilon}{g} \ ,
\end{align}

%-------------------
%------
\begin{figure}[t]
\centering
\includegraphics[width=3.5cm,pagebox=cropbox,clip]{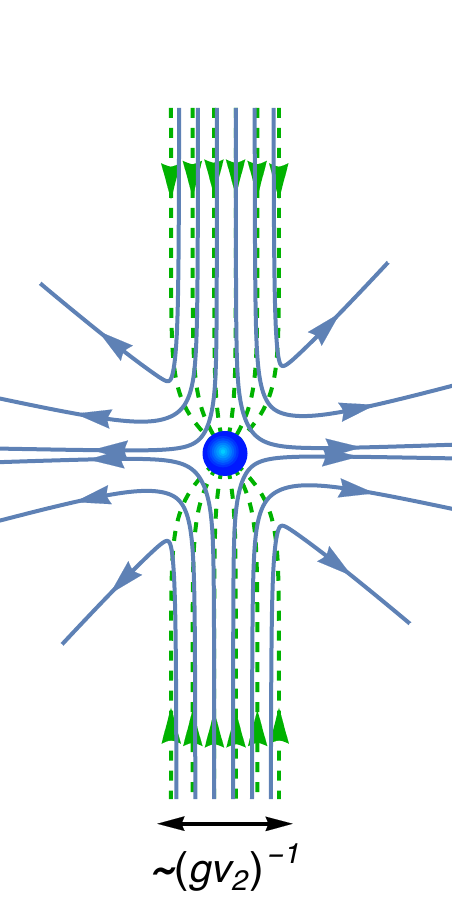}
\caption{\sl \small A schematic picture of the QED magnetic flux induced by the bead solution through the kinetic mixing term (blue solid lines).
The dark magnetic flux is shown by the green dashed lines.
In the QED magnetic flux follows the dark magnetic flux inside the strings.
The QED magnetic flux leaks out from
around the dark magnetic monopole, i.e., $|z|\sim (gv_2)^{-1}$, so that it satisfies the Bianchi identity of the U(1) gauge theory.
}
\label{fig:QEDflux}
\end{figure}
%-------------------

Now, the question is where the QED flux flowing into the region of $z \sim (gv_2)^{-1}$ goes.
The QED gauge field satisfies the Bianchi identity in the entire spacetime.
Therefore,
the QED magnetic flux cannot have no sources nor sinks.
Thus, the QED magnetic flux leaks out into the bulk space from the
region, $z\sim (gv_2)^{-1}$.
In Fig.\,\ref{fig:QEDflux}, we show a schematic picture to show
how the QED magnetic flux leaks out into the bulk.

%-------------------
%------
\begin{figure}[t]
\centering
\includegraphics[width=8.5cm,pagebox=cropbox,clip]{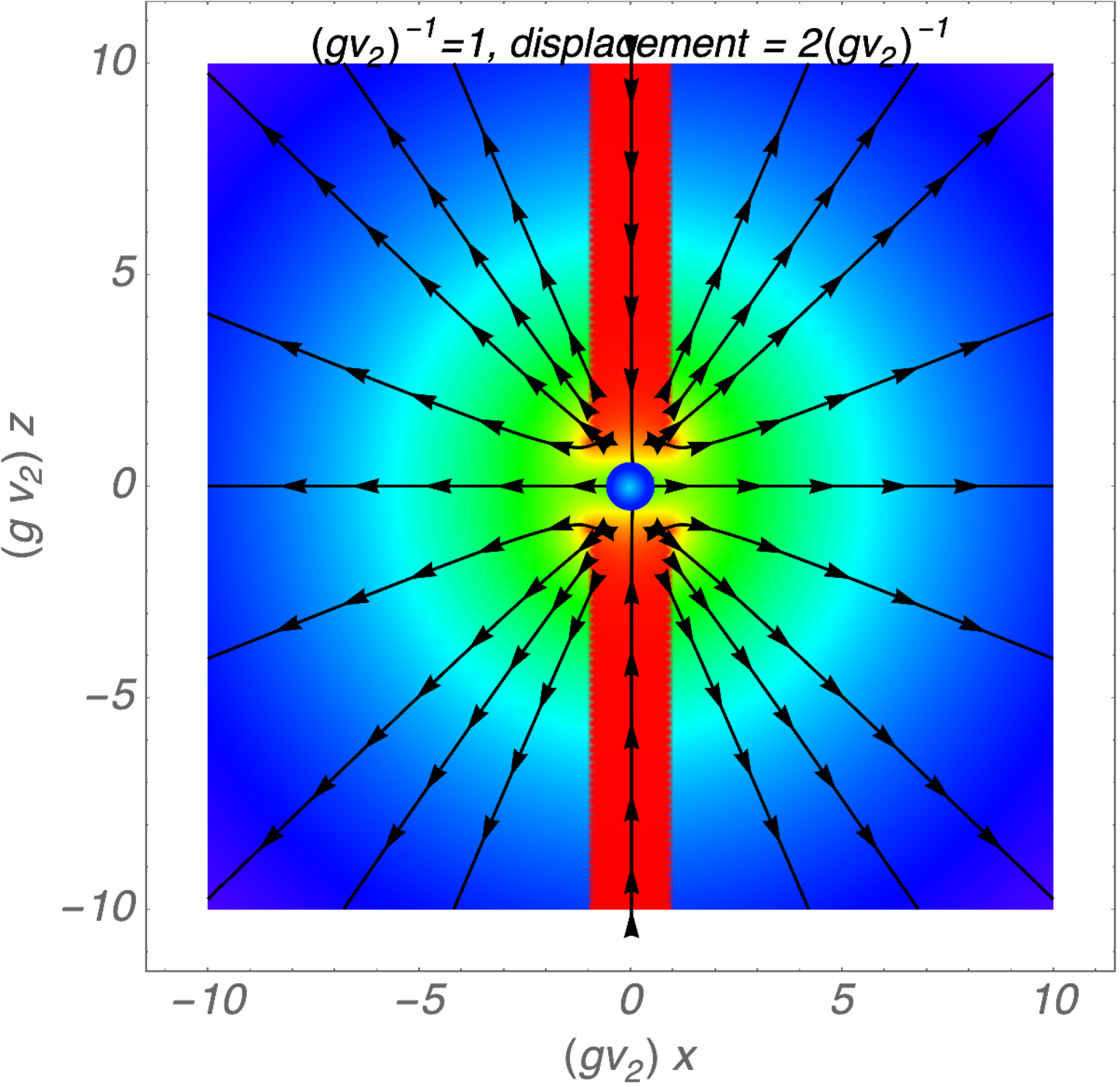}
\caption{\sl \small 
The QED magnetic flux made by the QED fluxes confined in the dark cosmic strings.
We approximate the configuration by the magnetic field made by two half infinite solenoids, which induces the spherical magnetic flux.
The color density shows the absolute strength of the QED magnetic field.
The strength decreases from red to blue in arbitrary unit.
The monopole-like flux extends from the center of figure is induced not by 
the dark monopole but by the dark cosmic strings.
}
\label{fig:QEDfield}
\end{figure}
%-------------------

Since no analytic expression of the bead solution has been known, it is  
difficult to solve the field equations
of the QED gauge field around it analytically.
However, we can gain insight of the
solution in the following way.
As we have discussed above, the dark (anti-)string
induces the QED magnetic flux along the (anti-)string at $|z|\gg (g v_1)^{-1}$.
Thus, the QED magnetic flux of this system can be well approximated by the magnetic flux made by two solenoids with opposite currents.%
\footnote{In the Appendix~\ref{sec:Solenoid}, we give the magnetic fields made by a finite solenoide in Ref.\,\cite{osti_4121210}.}
In Fig.\,\ref{fig:QEDfield}, we show 
the field lines of the QED magnetic flux on the $(x,z)$ plane made
by two solenoids with opposite circular currents.
Here,  the solenoids   
extend from 
$z = \pm (gv_2)^{-1}$ to $z\to \pm\infty$.
The radius of them is set to be
$(gv_2)^{-1}$.
The magnetic flux made by solenoids satisfy the Bianchi identity of QED automatically, and hence, the approximated solution using two half infinite solenoids captures the important property of the QED magnetic flux around the bead solution.
In the figure, we also show the strength of 
the QED magnetic flux by the color density.
The strength decreases 
from red to blue in arbitrary unit.
The figure shows that the leaked QED flux looks spherical viewed from $r \gg (g v_2)^{-1}$.

The figure shows that the magnetic flux of QED made by the two solenoids diverges spherically from the region $r\ll (gv_2)^{-1}$ viewed from a distance. The spherical flux is reasonable as the in-flowing flux to the monopole region is confined in $\rho \sim (g v_2)^{-1}$ and it leaks from a tiny region in $r \sim (g v_2)^{-1}$.
As a result,
we expect that the spherical QED magnetic flux is induced by the dark bead solution  through the kinetic mixing,
\begin{align}
    B_i = \frac{1}{2}\varepsilon_{ijk}F_{jk} \sim \frac{\epsilon}{g}\frac{x^i}{r^3}\ ,
\end{align}
which looks like a magnetic monopole of QED from a distance.%
\footnote{The corresponding magnetic field strength is $|B_i|\simeq (\epsilon/g)\,\mathrm{G}\times (76\mu\mathrm{m}/r)^{2}$ for $r\gg(gv_2)^{-1}$.}
The QED charged particles feel the Lorentz force around the dark magnetic monopole.
We call this configuration of QED gauge field, the pseudo monopole.

%--------
\begin{figure}[t]
\centering
\includegraphics[width=12cm,pagebox=cropbox,clip]{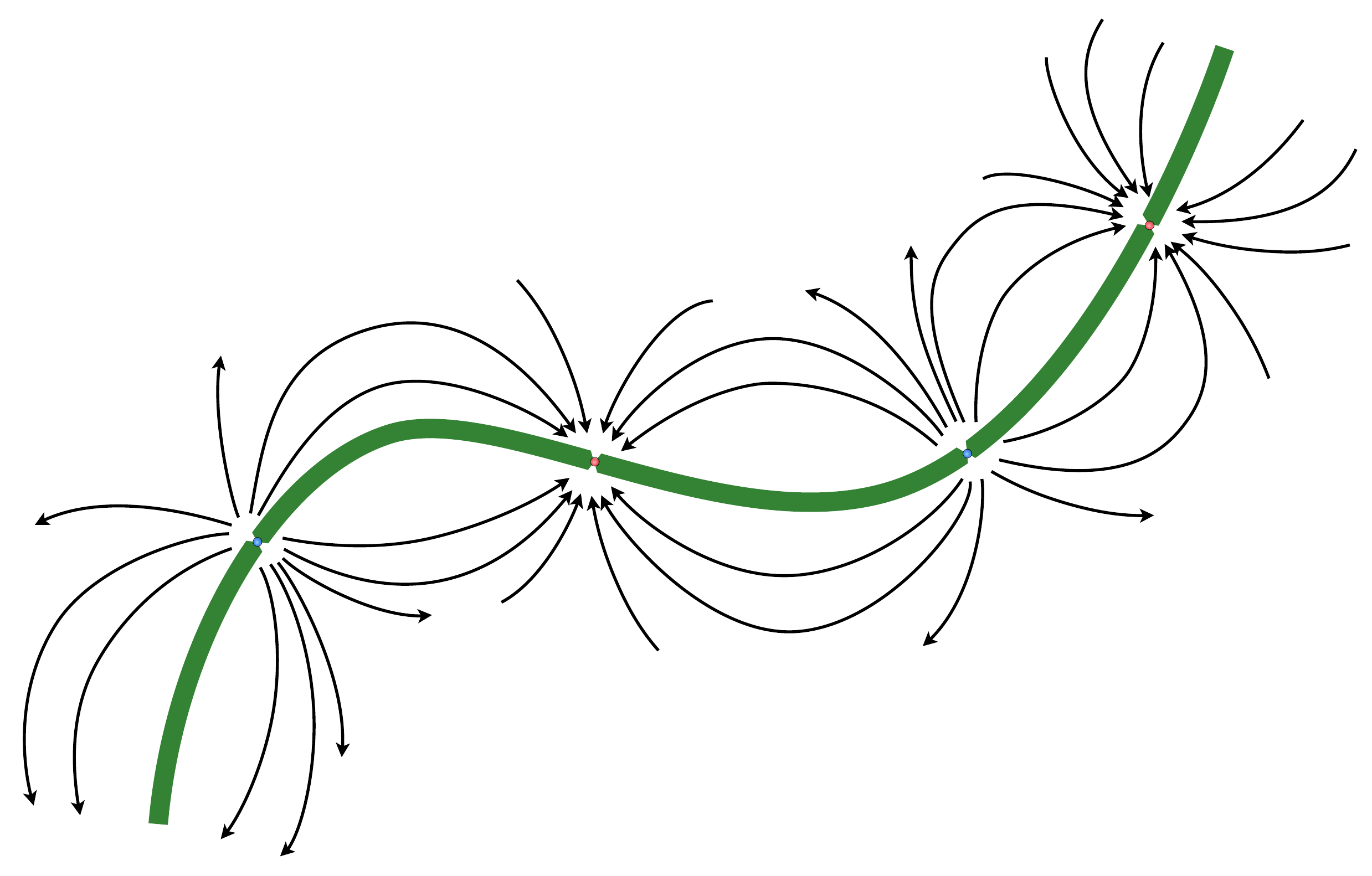}
\caption{\sl \small 
A schematic picture of the magnetic necklace.
The dark magnetic flux is trapped inside 
the necklace (the green line).
In the presence of the kinetic mixing,
the QED magnetic flux (the black lines) leaks out from the 
positions of the (anti-)monopole.
}
\label{fig:necklace}
\end{figure}
%-------------------

Finally, let us consider the network of the bead solution, the necklace.
When the dark monopoles appear at the first phase transition, it is expected that the total dark magnetic charge in the entire Universe
should be vanishing.
Thus, we expect that the same numbers of the dark monopoles and the dark anti-monopoles are formed.
At the second transition, the (anti)-dark monopoles are trapped into the dark cosmic string which form a network, i.e., the necklace.
In the necklace, the dark magnetic flux is confined.
In the presence of the kinetic mixing, the QED magnetic flux leaks from the positions of (anti)-monopoles, 
and hence, the dark necklace becomes the magnetic necklace of QED.
In Fig.\,\ref{fig:necklace}, we show 
a schematic picture of the magnetic necklace.

\section{Numerical Simulation}
\label{sec:simulation}
To confirm the formation of the necklace and the QED magnetic field induced through the kinetic mixing, we perform numerical simulations in a 3D expanding box. 
As discussed in the previous section, we introduce two adjoint scalar fields, $\phi_1$ and $\phi_2$ of SU(2)$_{\mathrm{D}}$.
The first scalar field, $\phi_1$, develops the monopoles at the first phase transition at $v_1$ and involves the kinetic mixing between U(1)$_{\mathrm{QED}}$ and SU(2)$_{\mathrm{D}}$ gauge fields as discussed in Sec.~\ref{sec:MonopoleEpsilon}. 
The second scalar field, $\phi_2$, then develops the cosmic strings at the second phase transition at $v_2\ll v_1$, which is expected to confine the U(1)$_{\mathrm{D}}$ magnetic field into the cosmic strings. The U(1)$_{\mathrm{QED}}$ magnetic field is induced by the U(1)$_{\mathrm{D}}$ magnetic field through the kinetic mixing.

The action is summarized as
\begin{align}
\mathcal{L} = -\frac{1}{4}F_{\mu\nu}F^{\mu\nu} -\frac{1}{4}F'{}^{a}_{\mu\nu}F'{}^{a\mu\nu} + \frac{\phi^a}{2\Lambda}F'{}^{a}_{\mu\nu}F^{\mu\nu}
+D_\mu\phi^a_1 D^\mu\phi^a_1
+D_\mu\phi^a_2 D^\mu\phi^a_2
-V(\phi_1,\phi_2),
\end{align}
where the potential is given by Eq.~(\ref{eq:potential}).
The governing equations for the scalar fields and the gauge fields are given by varying the action.
We take the temporal gauge, $A^a_0 = 0$, with the non-Abelian generalization of Gauss law as a constraint.

We assume the radiation dominated Universe as a cosmic background and that the scalar fields have thermal distributions at the initial time with the initial temperature $T = v_1$,%
\footnote{Our purpose of the present simulations is to demonstrate the development of the necklace and the confinement of the magnetic fields. Thus, the  background solution and the initial distribution of the scalar fields are of little importance. However, we have to take care of the initial conditions, if we would like to measure the physical properties of the necklace, e.g., the correlation length of the necklace, or to accelerate the relaxation of the fields in the computational box as in Ref.~\cite{Hindmarsh:2016dha}.} 
and impose the periodic boundary conditions spatially. 
We use the conformal time $\eta$ and set the initial condition at $\eta = 0$.
Without loss of generality, we fix the initial scale factor to be the unity, $a(0)=1$.

\begin{table}[!t]
    \centering
    \begin{tabular}{c|c|c}
        \hline\hline
                  & I & II \\ \hline
        grid size & 384 & 256 \\
        time step & 14400 & 25600 \\
        $s_{\rm i}$ & 60 & 30 \\
        $s_{\rm f}$ & 2 & 0.2 \\
        $a_{\rm f}$ & 30 & 150 \\ 
        $v_2/v_1$ & 0.3 & 0.3 \\
        $\lambda_{10}$ & 1 & 1 \\
        $\lambda_{20}$ & 1 & 1 \\
        $\kappa_0$ & 2 & 2 \\
        $\epsilon$ & 0.2 & 0.2 \\ 
        $g$ & $1/\sqrt{2}$ & 1 \\ \hline\hline
    \end{tabular}
    \caption{The model parameters for the numerical simulation. }
     %% t54a18 (candidate:t58a7)
    \label{tab:param}
\end{table}

We solve the field equations by the Leap-Frog scheme and the 2nd-order finite differences for the spatial derivatives. The model parameters are tabulated in Tab.\,\ref{tab:param}.
Following Ref.~\cite{Hindmarsh:2016dha},
 we employ the Press-Ryden-Spergel algorithm~\cite{Press:1989yh}
to maintain the width of strings and the size of monopoles in the comoving box to gain a wide dynamic range of the simulation.
In this algorithm, the coupling constants, $\lambda_i$ and $g$, scale as
\begin{align}
 \lambda_i(\eta) = \frac{\lambda_{i0}}{a^2}\ , \quad g(\eta) = \frac{g_0}{a}\ ,
 \label{eq:PRS}
\end{align}
with $\lambda_{i0}$ and $g_{0}$ being their initial values ($g$, $\lambda_{1,2}$ in Tab.\,\ref{tab:param}, respectively). 
We turn on this scaling at $a(\eta)=2$.

%-------------------
\begin{figure}[t]
{
\centering
\includegraphics[width=7.8cm,pagebox=cropbox,clip]{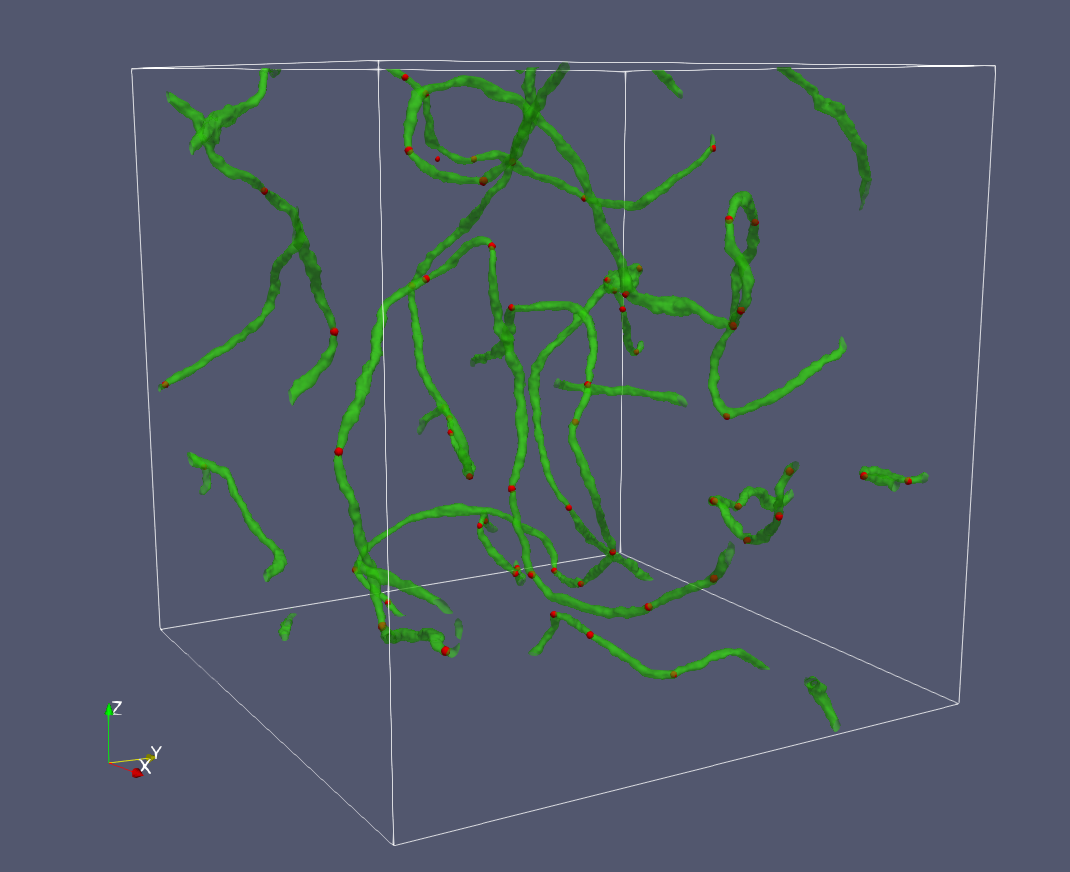}
}
\caption{\sl \small 
Cosmic beads network, i.e., the necklace. The red and green surfaces are the isosurface of
$|\phi_1|=0.5v_1$ and $|\phi_2|=0.06v_1$, respectively.
The figure shows that the magnetic monopoles (or the beads)
appearing as red points are connected by the cosmic strings.
}
\label{fig:numerical_necklace}
\end{figure}
%-------------------

In Fig.\,\ref{fig:numerical_necklace}, 
we show the time slice at the end of 
simulation with the parameter set~I  in Tab.\,\ref{tab:param}.
The red and green surfaces are the isosurface of $|\phi_1|=0.5v_1$ and $|\phi_2|=0.06v_1$, respectively.
This shows that a string (green) connects to two monopoles (red) at its both ends, or, in other words, two strings are connected by a monopole. 
This configuration confirms the formation of the beads solution as well as their network, the  necklace (see also Ref.~\cite{Hindmarsh:2016dha}).

At the early time of the simulation, when the energy density of the scalar fields become smaller than $\order{\lambda_1v_1^4}$, the SU(2)$_{\mathrm{D}}$ is broken to U(1)$_\mathrm{D}$, and then the monopoles are formed.
After a while when the scalar fields are well relaxed and their energy density becomes smaller than $\order{\lambda_2v_2^4}$, the residual U(1)$_\mathrm{D}$ is totally broken by $|\phi_2| \simeq v_2$, and then the cosmic strings are formed.
We define $s(\eta)=aL/(H^{-1})$ to characterize the box size where $L$ is the physical box size at the initial time and $H$ is the Hubble parameter.
We set the initial and final values of $s(\eta)$ to be $s_\mathrm{i}=60$ and $s_\mathrm{f}=2$ in this simulation for the set~I. Note that the choice of $s_\mathrm{f}=2$ ensures to suppress the unphysical effects arising from the finiteness of the periodic box on the necklace.

%-------------------
\begin{figure}[t]
{
\centering
\includegraphics[width=7.5cm,pagebox=cropbox,clip]{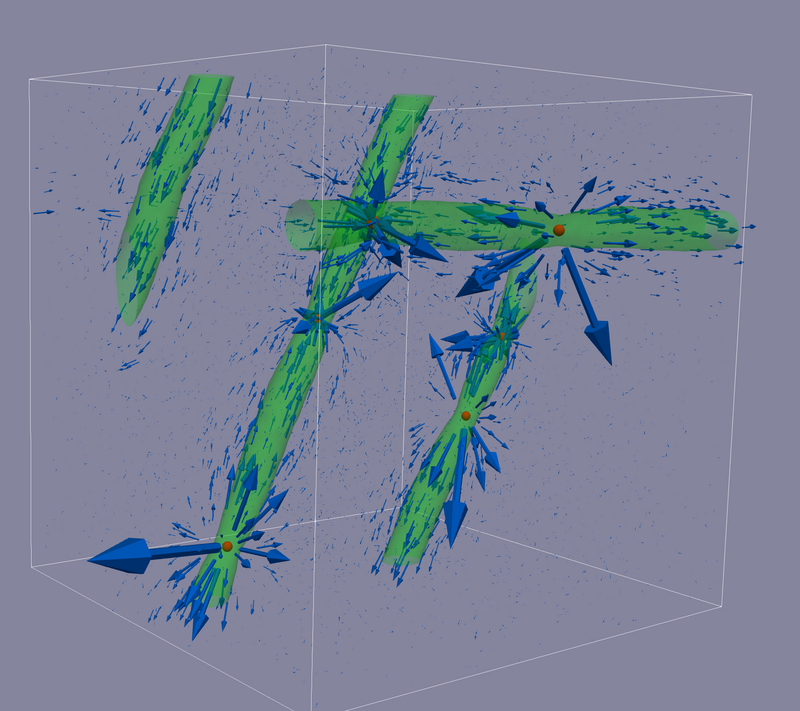}
}
\caption{\sl \small 
The snapshot of the necklace in the dark sector at the simulation end, $\eta v_1=79.0$, with the parameter set II  in Tab.\,\ref{tab:param}.
The red and green surfaces are the isosurface of
$|\phi_1|=v_1/2$ and $|\phi_2|=v_2/2$, respectively. The blue arrow represents the 
U(1)$_{D}$ magnetic field computed by Eq.\,\eqref{eq:effective_U1}.
The drawing points for the blue arrows are chosen randomly, so if the drawing points are close to the monopoles, the corresponding arrows are accidentally large.
This figure shows the magnetic fields around the strings are well aligned to them, while the configuration around the monopoles becomes like a hedgehog shape. 
}
\label{fig:confine}
\end{figure}
%-------------------

Next, we discuss how the magnetic flux from the monopoles are confined in to the strings.
The network simulation shown before yields wriggling strings with a number of kinks. 
Therefore, the magnetic field associated with them has a highly complicated configuration.
To avoid complexity, we perform a long-time simulation with a smaller box where we set $s_{\rm i}=30$ and $s_{\rm f}=0.2$.
In Fig.\,\ref{fig:confine}, we show 
the snapshot of the necklace at the simulation end
with the parameter set~II in Tab.\,\ref{tab:param}.
Thanks to the choice of $s_{\rm f}<2$, the long strings in the box feel like they lie in a 3 dimensional-torus, instead of $R^3$ space.
Therefore, the strings 
in Fig.\,\ref{fig:confine} are fully stretched by 
their tension.
Such a situation is not realistic in the real Universe, since our Universe may not be 3D-torus. 
However, our present purpose of this simulation is to confirm the confinement of the magnetic flux arising from monopoles into the strings. 
To see this as clearly as possible, it is convenient for the strings to be straight at the end of simulation.
Note that we show the isosurface with $|\phi_1|=v_1/2$ and $|\phi_2|=v_2/2=0.15v_1$ in Fig.\,\ref{fig:confine}, 
which is different from the choices for the 
network simulation in Fig.\,\ref{fig:numerical_necklace}. 
Accordingly, the size of the strings looks larger in comparison with those of the monopoles, since $v_2/v_1 = 0.3$.  

To quantify the gauge-invariant $U(1)_D$ magnetic field, 
we define the effective field strength proposed by Ref.~\cite{tHooft:1974kcl}, 
\begin{align}
 F^{\prime\mathrm{(eff)}}_{\mu\nu} = \frac{1}{|\phi|}\phi^a(\partial_\mu A^{\prime a}_\nu - \partial_\nu A^{\prime a}_\mu + g\epsilon^{abc}A^{\prime b}_\mu A^{\prime c}_\nu) - \frac{1}{g|\phi|^3}\epsilon_{abc}\phi^a D_\mu\phi^b D\
_\nu\phi^c\ .
\label{eq:effective_U1}
\end{align}
Here, the gauge coupling constant, $g$, depends on time as given in Eq.\,\eqref{eq:PRS}.
This effective field strength converges to Eq.\,\,\eqref{eq:PRS} for $r\to \infty$, 
while for, for example, $A^{1,2}_\mu=0, A^3_\mu\ne 0, \phi^{1,2}=0$ and $\phi^3\ne 0$, we obtain $F^{\rm (eff)}_{\mu\nu} =\partial_\mu A^3_\nu- \partial_\nu A^3_\mu$, which 
gives the usual field strength for U(1) gauge field. Thus, it is natural to define the effective magnetic field  as 
\begin{align*}
 B^{\prime \mathrm{(eff)}}_i = \frac{1}{2}\epsilon_{ijk}F^{\prime\mathrm{ (eff)}}_{jk},
\end{align*}
where $\epsilon_{ijk}$ is the antisymmetric tensor with $i,j,k=1,2,3$.

In previous sections, we have used the effective field strength $\mathcal{F}'_{\mu\nu}$ defined in Eq.\,\eqref{eq:effectiveF} instead of $F^{\prime\mathrm{(eff)}}_{\mu\nu}$. As discussed around Eq.\,\eqref{eq:bianchi}, the Bianchi identity, $\partial_\mu\tilde{\mathcal{F}}^{\prime\mu\nu}=0$, is satisfied only far from the monopole origin. On the other hand, for $F^{\prime\mathrm{(eff)}}_{\mu\nu}$, the usual Maxwell equations including the Bianchi identity,  $i.e.$ $\partial_\mu F^{\prime\mathrm{(eff)}\mu\nu}=0$ and $\epsilon^{\mu\nu\rho\sigma}\partial_\mu F^{\prime\mathrm{(eff)}}_{\rho\sigma}=0$, are obtained except where $\phi^a=0$. At $r\gg (gv)^{-1}$, the second term in Eq.\,\eqref{eq:effective_U1} is close to zero faster than the first term, which leads to $\mathcal{F}'_{\mu\nu}=F^{\prime\mathrm{(eff)}}_{\mu\nu}$  $(r\gg (gv)^{-1})$. Thus, we obtain the same magnetic charge of a monopole for both the field strengths while $\mathcal{F}'_{\mu\nu}$ has been used in Eq.\,\eqref{eq:Mcharge}.

In Fig.\,\ref{fig:confine}, we show $B^{\prime\mathrm{(eff)}}_{i}$ as blue arrows.
The drawing points for the blue arrows are chosen randomly, so if the drawing points are close to the monopoles, the corresponding arrows are accidentally large.
The figure shows that the magnetic flux is
aligned along the string and confined in the string as expected.

%-------------------
\begin{figure}[t]
{
\centering
\includegraphics[width=7.5cm,pagebox=cropbox,clip]{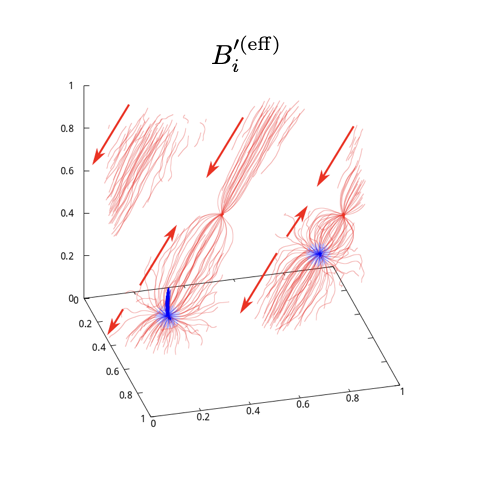}  
\includegraphics[width=7.5cm,pagebox=cropbox,clip]{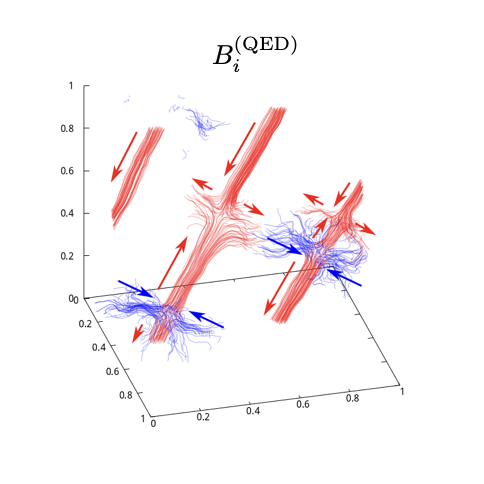}
}
\caption{\sl \small 
The stream line constructed by the magnetic field $B^{\prime\mathrm{(eff)}}_i$ ({\it left}) and $B^{(\mathrm{QED})}_i$ ({\it right}) 
at $\eta v_1=79.0$. We focus on the magnetic flux arising from 
two lower monopoles on the two diagonal cosmic strings in Fig.~\ref{fig:confine}.
The axis labels indicate the spatial coordinate normalised by the box size.
The stream lines start from a certain distance from the monopole points of the monopoles.
The red ones are the stream lines flowing out of the monopole region, while the blue ones are those for the ``negative" length parameter.
(See the Appendix~\ref{sec:stream} for details of the stream line.)
The arrows with a corresponding color indicate the direction of the flux.
In the left panel, the blue lines show that the magnetic flux are ending at the monopole points, while the red ones show that the magnetic flux flows along the cosmic strings and reach the left monopoles.
The right panel shows that the induced $B_i^{\mathrm{(QED)}}$ flowing out of the monopole regions (red) follows the cosmic strings, while it actually goes outside of the monopole region (blue).
}
\label{fig:stream}
\end{figure}
%-------------------

Finally, we visualise how the effective magnetic field, $B^{\prime\mathrm{(eff)}}_{i}$, is confined in a string.
For this purpose, we compute the stream lines associated with the magnetic field.
The stream lines (with the length parameter $\zeta$) start at certain distances from the monopole points.
In what follows, the red stream lines are those flowing out of the monopole region, while the blue ones are those for the ``inverse" length parameter.
That is, we solve the equation of the stream lines in Eq.\,\eqref{eq:stream} from $\zeta=0$ to $\zeta=0.075$ (red lines) and to $\zeta=-0.075$ (blue lines).
We also take the initial radius parameter $\alpha_*=3.3$ ($\simeq v_1/v_2$), which corresponds to 3.3 times the effective monopole size  
(see the  Appendix~\ref{sec:stream} for the details how to construct the stream lines). 
In the following, we focus on the two diagonal cosmic strings
in  Fig.\,\ref{fig:confine}, each of which has
a pair of monopole and anti-monopole.

In the left panel of Fig.\,\ref{fig:stream}, the blue lines
are localized around the two monopoles, which show that the magnetic flux are starting from the monopoles. 
Note that, due to the numerical error, the blue lines are overshot at the left-bottom monopole.
The red stream lines flow along the cosmic string, which show that the magnetic flux starting from the two lower monopoles are confined in the cosmic strings.
The figure also shows the convergence of the stream lines to another monopole in the each string. 
Thus, the left panel confirms the confinement of the magnetic flux in the bead solution.

We also compute the stream lines of the QED magnetic field induced through the kinetic mixing, 
\begin{align*}
 B^{\mathrm{(QED)}}_i = \frac{1}{2}\epsilon_{ijk}F_{jk},
\end{align*}
where $F_{jk}$ is the field strength of the U(1)$_\mathrm{QED}$ gauge field.
We focus on the magnetic flux from the same  lower monopoles discussed above, and solve the stream equation (\ref{eq:stream}) from $\zeta=0$ to $\zeta=0.4$ (red lines) and to $\zeta=-0.4$ (blue lines).
The figure shows that the red stream lines flow along the string, similar to the $B^{\prime\mathrm{(eff)}}_i$ sector.
The blue stream lines, however, are escaping  from the strings at the monopoles.
In other words, the QED magnetic flux induced on the cosmic string flowing into the monopole region leaks out to the bulk space.
Thus, the simulation also confirms the 
expected feature of the QED magnetic flux around the bead solution in Fig.\,\ref{fig:QEDflux}.

\section{Conclusion}
\label{sec:conslusion}
In this paper, we discussed how the topological defects 
in the dark sector affect the SM sector through the gauge kinetic mixing between the QED photon and the dark photon.
In particular, we considered the $\mathrm{SU}(2)_\mathrm{D}$ extension of the dark photon model with the breaking pattern $\mathrm{SU}(2)_\mathrm{D} \to \mathrm{U}(1)_\mathrm{D} \to \mathbb{Z}_2$.
In this model, the dark monopole is formed at the first phase transition
and the dark cosmic string is formed at the second phase transition.
As an interesting feature of the model, the dark monopole is trapped 
into the cosmic strings and forms
the bead solution.

We showed that the dark string induces a non-vanishing QED magnetic flux inside the dark string through the kinetic mixing. The dark monopole, on the other hand, does not induce the QED magnetic flux even 
in the presence of the kinetic mixing.
Finally, we found that the bead solution induces the QED magnetic flux which looks like the magnetic monopole viewed from a distance, 
which we call the pseudo monopole.
We also confirmed the formation of the pseudo monopole by the 3+1 dimensional numerical simulation.

In this paper, we have focused on the theoretical aspects of the dark topological defects in the presence of kinetic mixing.
Detailed studies of phenomenological, astrophysical, and cosmological implications will be given elsewhere.
Here, we only comment on the energy density of the necklace and the monopole in the Universe.
The numerical simulation of the cosmological evolution of the necklace
for $\epsilon = 0$ in Ref.~\cite{Hindmarsh:2016dha} suggests that the string solution follows the scaling solution, i.e., $\rho_{\mathrm{string}} \sim T_{\mathrm{str}} H^2$ where $T_\mathrm{str}$ is the string tension and $H$ is the Hubble parameter.
The monopole-to-string energy density ratio, on the other hand, decreases as the inverse of the scale factor (see also Ref.~\cite{BlancoPillado:2007zr}).
Thus, the dark defects do not contribute to the energy density of the Universe significantly, as long as the second phase transition takes place before the dark monopole dominates over the energy density of the Universe.

As a crude estimate, the monopole would dominate the energy density of the Universe at the temperature around $T_{\mathrm{dom}}\sim (M_{\mathrm{Pl}} r_c^2)^{-1}$, by approximating the annihilation cross-section of the monopole by its geometrical size, $r_c \sim (gv_1)^{-1}$.
Here $M_{\mathrm{Pl}} \simeq 2.4\times 10^{18}$\,GeV is the reduced Planck scale.
Thus, as long as the temperature of the second phase transition, $\order{v_2}$ is much larger than the monopole domination temperature, $(gv_1)^2/M_{\mathrm{Pl}}$, the energy density of the dark defects are expected to be subdominant.

\begin{acknowledgments}
This work is supported in part by JSPS KAKENHI Grant Nos. JP17H02878, JP18H05542 (M.I.), JP21K03559 (T.H.); World Premier International Research Center Initiative (WPI Initiative), MEXT, Japan (M.I.).
\end{acknowledgments}
\appendix
\section{SU(2) convention}
\label{sec:cov}
In the $2\times 2$ matrix representation of SU$(2)$, we define the covariant derivative 
of the adjoint representation is given by,
\begin{align}
\label{eq:covADJ}
    D_\mu \phi = \partial_\mu \phi - i g [A_\mu,  \phi] \ ,
\end{align}
where
\begin{align}
\phi = \phi^a\tau^a\ , \quad A_\mu = A_\mu^a \tau^a\ ,
\end{align}
 with $\tau^{a=1,2,3}$ being the half of the Pauli matrices.
Under the gauge transformation,
\begin{align}
    &\phi \to \phi' = \hat{g} \phi \hat{g}^\dagger \ ,\\
    &A_\mu \to A_\mu' = \hat{g} A_\mu \hat{g}^\dagger - \frac{i}{g}(\partial_\mu \hat{g})\hat{g}^\dagger\ ,
\end{align}
the covariant derivative transforms,
\begin{align}
     D_\mu \phi \to  \hat{g}(D_\mu \phi)\hat{g}^\dagger\ .
\end{align}

Let us also note that the U(1) gauge transformation corresponding to the $a=3$ rotation of SU(2). Under this U(1) transformation, the gauge field and the complex scalar transform,
\begin{align}
& A^3_\mu \to A^3_\mu  + \frac{1}{g} \partial_\mu \alpha\ , 
&(\phi^1 - i \phi^2) \to e^{i \alpha} (\phi^1 - i \phi^2)\ , 
\end{align}
which shows that $(\phi^1-i\phi^2)$ has the U(1) charge $+1$.
We can also check that the covariant derivative in Eq.\,\eqref{eq:covADJ} is reduced to
\begin{align}
    D_\mu \tilde{\phi} = (\partial_\mu - i g A_\mu)\tilde{\phi}\ ,
\end{align}
where $\tilde\phi = (\phi_1 - i \phi_2)/\sqrt{2}$.
\section{The Magnetic Field of a Finite Solenoid}
\label{sec:Solenoid}
Let us consider a finite solenoid
with a radius $a$ along the $z$-axis with the  
the surface current density 
\begin{align}
    J_\varphi = j \delta(\rho-a)\ ,
\end{align}
in $|z| < L/2$.
Here, we use the cylindrical coordinate $(\rho,\phi,z)$ and $R$ is the radius of the solenoid.
The magnetic field around the finite solenoid is given by~\cite{osti_4121210},%
\footnote{The expressions in Eqs.\,\eqref{eq:Brho} and \eqref{eq:Bz} are obtained by integrating Eqs.\,(6) and (8) of Ref.~\cite{osti_4121210}.
}
\begin{align} 
\label{eq:Brho}
&B_\rho = \frac{j}{2\pi }
    \sqrt{\frac{a}{\rho}}
    \left[\left(
    \frac{k^2-2}{k}K(k^2) + \frac{2}{k}E(k^2)\right)\right]^{\xi_+}_{\xi_-}\ , \\
\label{eq:Bz}
&B_z =  \frac{j}{4\pi }
\frac{1}{\sqrt{{a\rho}}}
     \left[\xi k\left(K(k^2) + \frac{a-\rho}{a+\rho}\Pi(
     \ell^2,k^2)\right)\right]^{\xi_+}_{\xi_-} \ , \\
     &B_\phi = 0\ ,
\end{align}
where 
\begin{align}
    \xi_\pm = z \pm \frac{L}{2}\ , \quad k^2 = \frac{4a\rho}{(a+\rho)^2+\xi^2}\ , \quad 
    \ell^2 = \frac{4a\rho}{(a+\rho)^2}\ .
\end{align}
In the above expression, we have used the complete elliptic integral of the first kind $K(m)$,
the complete elliptic integral $E(m)$, and the complete elliptic integral of the third kind $\Pi(n,m)$, respectively.

%% ===============================================================================================
\section{Stream Line}
\label{sec:stream}

To visualise how the magnetic flux spread around monopoles, we compute the stream line $\boldsymbol{x}_s(\zeta)$, where $\zeta$ is the length parameter characterising the stream line, by solving
\begin{align}
  \frac{d\boldsymbol{x}_s}{d\zeta} = \boldsymbol{B}^{(\rm eff)}_i(\boldsymbol{x}_s(\zeta)), \label{eq:stream}
\end{align}
from a point, $\boldsymbol{x}_s(0)=\boldsymbol{x}_0$, close to the monopoles.
Here, the bold characters denote the spatial vector.

\begin{figure}[t]
{
\centering
\includegraphics[width=5.5cm]{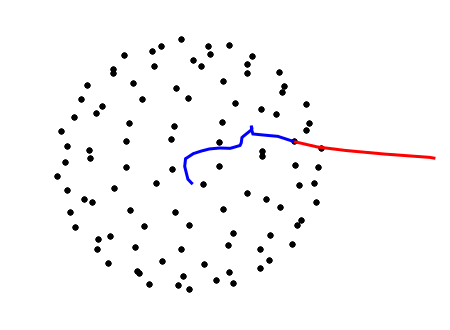}
\includegraphics[width=7.5cm]{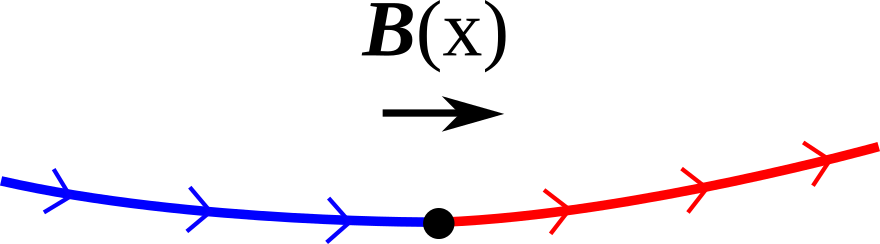}
}
\caption{\sl \small 
{\it Left} : Equally distributed points on a sphere enclosing a monopole. {\it Right} : Red curve is the stream line for positive $\zeta$ and blue one is that for negative $\zeta$. 
}
\label{fig:sample}
\end{figure}

Let us determine a set of the starting points, $\boldsymbol{x}_0$.
First we identify the volume, $V$, satisfying $|\phi_1|<v_1/2$ centred at a monopole core, and consider a sphere, whose radius is given as $r_* = \alpha_*(3V/4\pi)^{1/3}$ with positive constant $\alpha_*$, enclosing the monopole. Then we distibute $N_*$ points equally spaced on the sphere. We solve Eq.(\ref{eq:stream}) from these points. In the left panel of Fig.~\ref{fig:sample}, we show the equally distributed points on a sphere enclosing a monopole with $N_*=200$. Solving Eq.~(\ref{eq:stream} from $\zeta=0$ to $\zeta>0$, we obtain a red curve, whose tangential vector is equal to $\boldsymbol{B}(\boldsymbol{x})$,
and solving it to $\zeta<0$, we obtain a blue curve, as shown in the right panel of Fig.~\ref{fig:sample}.

%-------------------
%-------------------

Notice that the stream line, $\boldsymbol{x}_s(\zeta)$, constructed here 
is nothing but a line whose tangential vector is equal to $\boldsymbol{B}(\boldsymbol{x}_s(\zeta))$ at every point on the line, and thus 
does not represent the physical magnetic flux, $\Phi(\boldsymbol{x})$. By construction, the number of the stream line at the initial sphere is fixed and the number along the line is not proportional the the magnetic flux density thereat, whereas the number of the physical magnetic flux passing across a closed area is proportional to the magnetic flux density. However, the stream line can visualise how the magnetic field arising from a monopole can be confined in a string.

%\bibliographystyle{apsrev4-1}
%\bibliography{papers}

\end{document}